\documentclass[nofootinbib,10pt, twocolumn,prd,preprintnumbers,superscriptaddress,aps]{revtex4-2}

\usepackage{amsfonts}
\usepackage{graphicx}
\usepackage[colorlinks=true,linkcolor=blue,citecolor=teal,urlcolor=blue]{hyperref}
\usepackage{xcolor}
\usepackage{amsmath}
\usepackage{cases}
\usepackage{braket}
\usepackage[T1]{fontenc}
\usepackage{mathrsfs}
\usepackage{enumerate}
\usepackage{bm}
\usepackage{multirow}
\usepackage{comment}
\usepackage{makecell}
\usepackage{subfigure}
\renewcommand{\arraystretch}{1.5}

\newcommand{\til}{~}

\begin{document}

\title{Probing Interacting Dark Sector with the next generation of gravitational-wave detectors}

\author{Matteo Califano}
\email{matteo.califano@unina.it}
\affiliation{Scuola Superiore Meridionale, Largo San Marcellino 10, I-80138, Napoli, Italy}
\affiliation{INFN Sezione di Napoli, Compl. Univ. di
Monte S. Angelo, Edificio G, Via Cinthia, I-80126, Napoli, Italy}

\author{Ivan De Martino}
\email{ivan.demartino@usal.es}
\affiliation{{Departamento de F\'isica Fundamental and IUFFyM, Universidad de Salamanca,Plaza de la Merced, s/n, E-37008 Salamanca, Spain}}

\author{Ruth Lazkoz}
\email{ruth.lazkoz@ehu.eus}
\affiliation{Euskal Herriko Unibertsitatea UPV/EHU, 644 Posta Kutxatila, 48080 Bilbao, Spain}
\affiliation{EHU Quantum Center, University of the Basque Country UPV/EHU, PO Box 644, 48080 Bilbao, Spain}

\begin{abstract}
We have probed the capability of third-generation Gravitational Waves (GW) interferometers, such as the Einstein Telescope and Cosmic Explorer, to constrain a cosmological model with an interacting dark sector. We focused on GW events with a detected electromagnetic counterpart being the $\gamma$ or X emission of Gamma-Ray Burst, and a Kilonova emission. We assume the first one to be detected by the THESEUS satellite, while the second one to be detected by the Vera Rubin Observatory. We probed three different interaction kernels and found that the posterior estimation of the cosmological parameters is biased due to the existing degeneracies between the dark and matter sectors. { We also found that introducing an external prior on the matter density parameter breaks the degeneracy, removes the bias results, and improves the accuracy on the dark sector parameters.}
\end{abstract}
\preprint{ET-0472A-24}

\maketitle

\section{Introduction}\label{sec:intro}
The detection of Gravitational Waves (GWs) resulting from the merging of Binary Black Holes (BBH) and Binary Neutron Stars (BNS)\til\cite{Abbott2016,GW170817} opened a new groundbreaking avenue to explore the Universe probing the cosmological model and the underlying theory of gravity\til\cite{GW170817,Abbott2016b,Ezquiaga:2017ekz}. The gravitational waveform of BBH and, in particular, of BNS mergers may provide direct insights into the luminosity distance of the sources, promoting those systems, commonly referred to as \emph{standard sirens}\til\cite{Schutz:1986gp,Holz:2005df}, to rulers for measuring distances in the Universe. Standard sirens offer a complementary approach to the \emph{standard candles} for measuring distances because they do not require calibration from nearby sources to determine their luminosity distance, as for Cepheids and Supernovae Type Ia (SNeIa).

While offering an alternative method to determine distances in cosmology without the calibration issues of traditional methods, GWs present unique challenges. Their waveform contains crucial information about the systems, such as masses, spin, inclination angle, and distance. The latter, in turn, depends on the underlying cosmological model. Moreover, the information on masses and redshift is fully degenerate, and one needs prior knowledge of redshift from an electromagnetic counterpart to break such a degeneracy. 
For instance, if one is capable of unambiguously associating a host galaxy to the GWs event, then the redshift of the host galaxy can be used to break the degeneracy with the mass of the system, thus allowing to correctly estimate the distance\til\cite{Holz:2005df,Dalal:2006qt,Nissanke:2009kt,LIGO_H0_2017}. While we may effectively associate the host galaxy to the GWs event at low redshift, we cannot do the same at high redshift ($z\gtrsim 1$)\til\cite{DelPozzo:2011vcw,Muttoni:2023prw,Gray:2019ksv,Gray:2021sew,Mastrogiovanni:2023emh}. Alternatively, the simultaneous detection of a Gamma Ray Burst (GRB) emitted during the merger of a BNS system may serve to get the redshift \cite{GW170817}. In the literature, such cases are usually called {\em bright sirens}.

On the other hand, when an electromagnetic counterpart is not observed, one usually talks about {\em dark sirens}. In such cases, alternative methods can be employed to determine the redshift of the sources. 
These methods encompass translating the redshifted mass distribution to the source mass distribution through the assumption of some astrophysical principles such as the pair-instability supernovae (PISN) process\til\cite{Bond:1984sn,Mastrogiovanni:2021wsd,Mastrogiovanni:2023emh}, extracting redshift data from galaxy catalogs\til\cite{Schutz:1986gp,DelPozzo:2011vcw,Nishizawa:2016ood}, analyzing spatial clustering patterns between GW sources and galaxies\til\cite{Mukherjee:2019wcg,Mukherjee:2020hyn}, leveraging external insights into the distribution of source redshifts\til\cite{Ding:2018zrk,Ye:2021klk}, and using the tidal distortions of neutron stars\til\cite{Messenger:2011gi,Chatterjee:2021xrm,Ghosh:2022muc,Ghosh:2024cwc}.
Having bright or dark sirens translates into our capability of constraining cosmological parameters. For example, to achieve 1\% accuracy in the Hubble constant with next-generation gravitational wave detectors, one needs to detect $\sim100$ GRB associated with GW events spanning a redshift range of up to $\sim3$\til\cite{Califano:2022cmo}.

Nowadays, the third Gravitational-Wave Transient Catalog (GWTC-3) of the LIGO/Virgo/KAGRA (LVK) Collaboration\til\cite{KAGRA:2021vkt} provides three estimations of the Hubble constant adopting three distinct methodologies.
The first estimation of the Hubble constant is derived from the coincident detection of GRB 170817A with the GW event GW170817\til\cite{GW170817}. The LVK Collaboration estimated the redshift from the electromagnetic counterpart and obtained the first estimation of the Hubble constant using GW: $H_0= 70_{-8}^{+12}$ km s$^{-1}$ Mpc$^{-1}$ at a 68\% confidence level\til\cite{LIGO_H0_2017}. 
Subsequently, statistically associating a host galaxy to the 47 events of GWTC-3 with a signal-to-noise ratio (SNR) greater than 11, the LVK Collaboration constrained the Hubble constant to $H_0= 68_{-6}^{+8}$ km s$^{-1}$ Mpc$^{-1}$\til\cite{LIGO_H0_2021}. Finally, with the same data set but modeling the population distribution of intrinsic parameters of the GW sources (masses and spins), along with merger rate evolution, the LVK Collaboration bounded the Hubble constant to $H_0= 68_{-7}^{+12}$ km s$^{-1}$ Mpc$^{-1}$\til\cite{LIGO_H0_2021}.
Nevertheless, due to their low accuracy, these new Hubble constant measurements do not help to solve the so-called "Hubble Tension" between late-time and early-time estimations of $H_0$ in the framework of the {\em concordance} $\Lambda$ Cold Dark  Matter ($\Lambda$CDM) model\til\cite{Verde:2019ivm}. Indeed, $H_0$ is bounded with a statistical uncertainty of  $\sim10\%$ which is a factor 10 higher than the accuracy achievable by fitting the Cosmic Microwave Background (CMB) power spectrum\til\cite{Planck:2018vyg} and using Cepheids as late-time standard candles\til\cite{Riess:2019cxk}.

This scenario will dramatically change with the third generation (3G) interferometers: the Einstein Telescope (ET) in Europe\til\cite{Punturo:2010zz} and the Cosmic Explorer (CE) in the USA\til\cite{Evans:2021gyd}. The scientific potential of these 3G detectors is enormous, with anticipated observations of $10^5-10^6$ GW signals from compact binary coalescences over a few years of observations. ET and CE are poised to make groundbreaking discoveries in various disciplines, including astrophysics, cosmology, and fundamental physics\til\cite{Maggiore:2019uih,Branchesi:2023mws,Califano:2022syd,Califano:2023fbq,Califano:2023aji,DAgostino:2019hvh,DAgostino:2022tdk}. 
For example, the coincident detection of the GW signal with a GRB or a Kilonova (KN) emission may yield
the precision of $H_0$ below 1\% with only $\mathcal{O}(100)$ observations of multi-messenger BNSs\til\cite{Branchesi:2023mws}.
The higher number of combined events is also related to the observation of an electromagnetic counterpart from the forthcoming satellites and telescopes\til\cite{Ronchini:2022gwk}. For instance, to look for the coincident GRBs one could use the Transient High Energy Sources and Early Universe Surveyor (THESEUS), which will be able to detect the GRB up to redshift $\sim 8$\til\cite{THESEUS:2017wvz} and will allow achieving the 1\% accuracy on $H_0$ in 5 years of observations \cite{Califano:2022cmo}. Additionally, KN emission could be detected with the Vera Rubin Observatory (VRO)\til\cite{LSST:2008ijt,LSST:2018bbx,LSSTTransient:2018dvm} up to redshift $\sim  0.3$ and will allow achieving the $\sim5$\% accuracy on $H_0$ in 1 year\til\cite{Branchesi:2023mws}. 

However, our capability of getting the Hubble constant from the GW event with such a low accuracy is also related to the underlying cosmological model we want to test. Indeed, if one
extends the analysis to dark energy models beyond the simplest cosmological constant then one finds that our capability of achieving the $1\%$ accuracy on $H_0$ is significantly degraded \cite{Califano:2022syd}. Nevertheless, studying dark energy models is one of the primary scientific goals of ET \cite{Maggiore:2019uih}. Therefore, here we want to forecast the precision down to which the 3G detector of GWs will constrain cosmological models with an interacting dark sector.

There is no fundamental theory that dictates a specific coupling in the dark sector, making any proposed coupling model inherently phenomenological. However, some models may have a stronger physical justification than others. Various models of energy exchange have been considered, including simple functional forms in which the interaction kernel depends only on the scale factor. However, these models are incomplete because they cannot be fully tested against observations. After all, it is not possible to study how the cosmological perturbations behave. A satisfactory model must express the interaction in terms of, for instance, energy densities, the Hubble factor or other scalars built with covariant quantities to ensure thorough observational testing.

In Sec.\til\ref{sec:models},  we will briefly introduce the interaction models. In Sect.\til\ref{sec:mockdata}, we will summarize the procedure adopted to build mock data that will mimic the observations of ET and CE together with THESEUS and VRO. In Sect.\til\ref{sec:results}, we will give the main features of our statistical analysis and show the results. Finally, in Sect.\til\ref{sec:conclusion} we will give our final discussion and conclusions.

\section{Theoretical models}\label{sec:models}
This line of inquiry concerned with interactions between dark matter and energy has been progressing for more than three decades thanks to diverse motivations. One of the most explored ones was solving the coincidence problem  \cite{Chimento:2005xa,Zimdahl:2004hk}

We assume that the standard Friedmann equations are satisfied, and the cosmological model we will study represents a spatially flat universe fueled by dark matter, dark energy, and baryons. Therefore, the Friedmann equations read:
\begin{eqnarray}
H^2=\frac{8\pi G}{3}\left({\rho_{\rm dm}+\rho_{\rm de}+\rho_b,}\right)\,,\\
2\dot H=-(\rho_{\rm dm}+\rho_b+(1+w)\rho_{\rm de}).
\end{eqnarray}
In this framework, the luminosity distance is defined as
\begin{equation}\label{eq: luminosty_distance general}
    d_{L}(z) = 
    \frac{c(1+z)}{H_0} \int_{0}^{z}\frac{dz'}{E(z')}\,,
\end{equation}
where $z$ is the redshift, $c$ is the speed of light, $H_0$ is the Hubble constant, and $E(z)$ for the $\Lambda$CDM model is given by
\begin{equation}\label{eq: E_z LCDM}
    E^{2}(z)=\Omega_{\rm m,0}(1+z)^{3} + \Omega_{\Lambda,0}\,,
\end{equation}
while for more general dark energy models can be rewritten 
\begin{equation}\label{eq: E_z O(z)CDM}
    E^{2}(z)=\Omega_{\rm m,0}(1+z)^{3} + \Omega_{\rm de}(z)\,,
\end{equation}
where $\Omega_{\rm m,0}$ is the value  of the matter (dark matter + baryons) density parameter at $z=0$, and $\Omega_{\Lambda,0}$ the critical cosmological constant density and  $\Omega_{\rm de}(z)$ is the dark energy (DE) density parameter.

In interacting models, the total energy density of the dark sector is conserved, while the densities of dark matter and dark energy evolve as:
\begin{eqnarray}
\dot \rho_{\rm de}+3H\rho_{\rm de}(1+w)=-Q, \label{eq:interaction1}\\
\dot \rho_{\rm dm}+3H\rho_{\rm dm}=Q\label{eq:interaction2},
\end{eqnarray}
where $Q$ is the interaction kernel.  This is a derivation of an energy balance expressed in a covariant form as 
\begin{eqnarray}
\nabla_{\nu}{T_{\rm dm}^{\mu\nu}}=Q^{\nu},\,
\nabla_{\nu}{T_{\rm de}^{\mu\nu}}=-Q^{\nu}.
\end{eqnarray}
where $T^{\mu\nu}$ has the standard meaning of an energy-momentum tensor and $Q^{\mu}$ is the transfer four-vector.
In general, though, it is not straightforward to propose a form of this vector that leads to tractable or identifiable 
background equations such as (\ref{eq:interaction1}) and (\ref{eq:interaction2}). For that reason the preliminary step 
is often waived and the attention is shifted to the background dynamics, although an interest in cosmological perturbations would make it necessary to address the specification of $Q^{\mu}$, as discussed in the thorough analysis in \cite{Valiviita:2008iv}.

To circumvent the possible ``fifth-force" strong constraints, the baryonic sector is conserved on its own, so that the customary
\begin{eqnarray}
\dot \rho_b+3H\rho_b=0
\end{eqnarray}
equation holds and the same assumption is made for photons and neutrinos.
In this early stage,  for convenience, we write the interaction kernel compactly as $Q=H\Gamma$, where we have introduced an auxiliary function: $\Gamma\equiv\Gamma(\rho_{\rm dm},\rho_{\rm de})$. 
One of the main reasons why this coupling has been on high demand is its
mathematical simplicity, which leads to phase space portraits with the same number of dimensions (two) as the uncoupled
case. Moreover, as will become manifest later, this type of exchange allows for explicit final expressions of $H$ as a function of $a$ in many cases of interest, and this makes the scheme particularly interesting for observational constraint analyses. But, for the sake of fairness, the debate remains open as to whether it is physically reasonable to let the interaction depend on a global quantity such as $H$ instead of just local ones ($\rho_{\rm dm}$,$\rho_{\rm de}$). Nevertheless, from an indulgent frame of mind $H$ can be associated with a temperature dependence in the interaction (see again \cite{Valiviita:2008iv}).

More specifically, it is customary in the literature\til\cite{Chimento:2012kg,Wang:2016lxa,Wang:2024vmw} to set the function $\Gamma$ either proportional to the energy density of one of the two not conserved species or to a linear combination of them.

Here, we focus on a fairly general kernel with those specifications, namely:
\begin{equation}
    Q=H(\delta \rho_{\rm de}+
    \eta \rho_{m}).
\end{equation}
It is often overlooked that this type of interaction was first presented in \cite{Chimento:2012kg,Wang:2016lxa,Wang:2024vmw,Setare:2007we,Olivares:2006jr,Sadjadi:2006qp,Olivares:2006jr}, sparking significant activity in the field
until the later detailed observational analysis (of the $\eta=0$) case in both \cite{Gavela:2009cy, Yang:2022csz} gave the proposal a definitive boost. In this sense, the literature suggests that should the interaction be considered from a covariant point of view then phenomenological proposals in the form $Q^{\nu}= H(\delta \rho_{\rm de}u_{\rm de}^{\nu}+
    \eta \rho_{m}u_{m}^{\nu})$ could be appropriate.

In a broad sense, formulations have served as optimal arenas to show that the interaction does not necessarily preclude the presence of accelerated attractors \cite{vanderWesthuizen:2023hcl}.
Note that the dependence of $Q$ on $a$ remains implicit until the final $\rho_{\rm dm}$ and $\rho_{\rm de}$ are deduced. The interaction between the two components may modify the effective values of their equation of state parameters. 
Thus, equations \eqref{eq:interaction1} and \eqref{eq:interaction2} may be recast as
\begin{eqnarray}
\dot \rho_{\rm de}+3H\rho_{\rm de}(1+w+\delta)=-H\eta\rho_{\rm dm}\,,\\
\dot \rho_{\rm dm}+3H\rho_{\rm dm}(1-\eta)=H\delta\rho_{\rm de}\,.
\end{eqnarray}
If the interaction is not strong enough to completely spoil some of the main characteristics of the model, one expects dark energy dominance at early times, that is, $\rho_{\rm de}\gg\rho_{\rm dm}$ and effectively we will
have $w_{\rm de}^{\rm eff}=w+\delta$. Conversely if, at early times, dark matter will be the governor of the evolution then $\rho_{\rm de}\ll\rho_{\rm dm}$ and correspondingly $w_{m}^{\rm eff}=-\eta$. 

We aim to review known results in this front leading to exact parametrizations of $E(z)$ as a function of the scale factor $a$ or the redshift $z$. Note that having such expressions makes it rather unnecessary to examine these scenarios from a dynamical systems perspective. Discerning the stages of evolution is not complicated, as will become clear later. On the other hand, the dynamical systems approach becomes handy in other known cases. One of them is the case of an interaction kernel $Q$ that is not linear on $\rho_{\rm dm}, \rho_{\rm de}$ or $H$, and another possible one would be the presence of an explicit dependence on $a$ in $Q$ or the dark energy equation of the state parameter $w(a)$.

To be able to bring together the discussion of the two dark species whenever possible we define the density of the dark sector
\begin{equation}
   \rho_{\rm ds}=\rho_{\rm de}+\rho_{\rm dm}.
\end{equation}

We will carry a very general discussion following\til\cite{Chimento:2005xa} which is nevertheless best suited for the particular interaction of our choice. 

The separate conservation equations can be written as
\begin{eqnarray}
a \rho_{\rm dm}'+3\rho_{\rm dm}=\Gamma, \label{eq:rhomprime}\\
a \rho_{\rm de}'+3\rho_{\rm de}(1+w)=-\Gamma,
\label{eq:rhodeprime}
\end{eqnarray}
where the prime denotes differentiation with respect to the scale factor $a$ and we have chosen a constant equation of state parameter $w$. Note the disappearance of the dependence on $H$ due to this choice of variable and the specifications of $Q$ itself. Other choices lead to similar simplifications, such as differentiating with respect to $\log a^3$. This was a popular option when addressing these problems from a dynamical systems perspective. However, this would make it difficult to establish a direct comparison with old-school works in the area which are worth revisiting, such as \cite{Chimento:2011pk,Chimento:2012kg}.

With our preferred choice it is straightforward to find
\begin{eqnarray}
a \rho_{\rm ds}'+3(\rho_{\rm ds}+w\rho_{\rm de})=0, \label{eq:rhoprime}
\end{eqnarray}
and to use it to infer
\begin{eqnarray}
&&\rho_{\rm dm}=\frac{a\rho_{\rm ds}'+3(1+w)\rho_{\rm ds}}{3w},
\label{eq:rhom}\\
&&\rho_{\rm de}=-\frac{a\rho_{\rm ds}'+3\rho_{\rm ds}}{3w}\label{eq:rhode}.
\end{eqnarray}
Now differentiating equation \eqref{eq:rhoprime},
using equation \eqref{eq:rhodeprime} at a first step  and then equation \eqref{eq:rhode} at a second step, we arrive at
\begin{equation}
a^2 \rho_{\rm ds}''+(7+3w) a\rho_{\rm ds}'+{9 (1+w)
   }{\rho}_{\rm ds}={3  w}\Gamma \,.\label{rhoprimeprime}
   \end{equation}
   
Alternatively, we can replace $\rho_{\rm de}$ with $\rho_{\rm ds}-\rho_{\rm dm}$ in equation \eqref{eq:rhoprime} and differentiate it. The next step involves first applying equation \eqref{eq:rhom}, followed by equation \eqref{eq:rhom}.
 This will again produce equation \eqref{rhoprimeprime} unequivocally, proving the correctness of the procedure. 

Our final result concerning the specific form of the total energy density for the model we are constructing can be presented in a reasonably compact way if we introduce the  definition
\begin{equation}
    w_{\star}=\sqrt{(\delta -\eta )^2+w^2+2 w (\delta +\eta )},
\end{equation}
which lets us deduce that
  \begin{eqnarray}
\rho_{\rm ds}=\left(c_2 a^{3 w_{\star}}+c_1\right) a^{-\frac{3}{2} (\delta -\eta +w+w_{\star}+2)}\,.
   \end{eqnarray}

Now we have to replace the integration constants $c_1$ and $c_2$ with hopefully familiar parameters which will allow us to grasp the physical meaning of the results. To this aim, we will consider three different cases. 

\paragraph*{\bf CASE \protect\hypertarget{case I}{I}: $\mathbf{Q=\boldsymbol{\delta} H\boldsymbol{\rho}_{\rm de}}$.}
When the interaction is proportional to the DE energy density, the function $E^2(z)$ becomes:
\begin{equation}\label{eq: E(z) int DE}
E^2(z)=\Omega_{\rm m,0}(1+z)^3 + \frac{\Omega_{\Lambda,0}}{w_{\text{eff}}} (1+z)^{3} (-\delta + w(1+z)^{3w_{\text{eff}}})\,,
\end{equation}
where $\omega_{\text{eff}}=\omega-\delta$. It is worth noticing that the $\Lambda$CDM is recovered by setting $\omega= -1$ and $\delta = 0$.

\paragraph*{\bf CASE \protect\hypertarget{case II}{II}: $\mathbf{Q=\boldsymbol{\eta} H\boldsymbol{\rho}_{\rm dm}}$.}\hypertarget{case II}{}
In the case of  an interaction proportional to the DM energy density, the function $E^2(z)$ becomes
\begin{align}
    E^2(z)&=\Omega_{b,0}(1+z)^3 + \Omega_{\Lambda,0}(1+z)^{3(1+w)} \nonumber \\
    &+ \frac{\Omega_{\rm dm,0}}{w - \eta}(1+z)^{3}(-\eta (1+z)^{3w} + w(1+z)^{3\eta})\,.
\label{eq:E(z) int DM}
\end{align}
Let us notice that $\Lambda$CDM is recovered by setting $\omega= -1$ and $\eta = 0$.

\paragraph*{\bf CASE \protect\hypertarget{case III}{III}: $\mathbf{Q=\boldsymbol{\eta}H\left(\boldsymbol{\rho}_{\rm dm}+\boldsymbol{\rho}_{\rm de}\right)}$.}\hypertarget{case III}{}
If the interaction is related to the total dark energy densities (DM + DE) the function $E^2(z)$ can be recast as
\begin{align}
    E^2(z)&=\Omega_{b,0} (1 + z)^3 \nonumber\\
    &+ \Omega_{\rm dm,0} (1 + z)^3 \left( \frac{w (1 + z)^{3w - 3\eta}}{w - \eta} - \frac{(1 + z)^{3\eta} \eta}{w - \eta} \right)\nonumber\\
    &+ \Omega_{\Lambda,0} (1 + z)^3 \left( \frac{w (1 + z)^{3\eta}}{w - \eta} - \frac{(1 + z)^{3w - 3\eta} \eta}{w - \eta} \right)\,,
\label{eq:E(z) int total}
\end{align}
The $\Lambda$CDM cosmology is recovered by setting $\omega= -1$ and $\eta = 0$.

\section{Mock Data}\label{sec:mockdata}

To test the capability of the 3G GW detectors, the starting point is to build up the mock catalogs. Following\til\cite{Califano:2022cmo, Califano:2022syd, Califano:2023fbq}, we assign the redshift to GW sources sampling it from the following probability distribution\til\cite{Regimbau:2012ir,Cai:2016sby}
\begin{equation}\label{rate:unit_of_redshift}
    p(z) = \mathcal{N}\frac{R_m (z)}{1+z}\frac{dV(z)}{dz},
\end{equation} 
where $\mathcal{N}$  and $dV(z)/dz$ are the normalization factor and the comoving volume element, respectively. Moreover, $R_m (z)$ is the merger rate per unit of volume in the source frame and takes the form\til\cite{Regimbau:2009rk,Regimbau:2016ike,Meacher:2015rex}
\begin{equation}\label{merger_rate}
    R_{m} (z) = R_{\rm m,0} \int_{t_{min}}^{t_{max}} R_f[t(z)-t_d] P(t_d) d t_d \ ,
\end{equation}
where $R_f[t(z)-t_d]$ is the Star Formation Rate (SFR), and $P(t_d)$ is the time delay distribution. Based on population synthesis models\til\cite{Lipunov:1995ct,deFreitasPacheco:2005ub,Belczynski:2006br}, the time delay distribution is set the power law
\begin{equation}
   P(t_d)\propto t_{d}^{-1} \ ,
\end{equation}
while the SFR can be parametrized using the Madau-Dickinson model\til\cite{Madau:2014bja} as follows:
\begin{equation}
    R_f(z)=\left[1+(1+z_p)^{-\gamma-\kappa}\right]\frac{(1+z)^{\gamma}}{1+\left(\frac{1+z}{1+z_p}\right)^{\gamma+\kappa}}\,,
\end{equation}
where the parameters of the SFR are $\gamma = 2.6$, $\kappa = 3.1$ and $z_p=2$.  We set the integration extremes of equation\til\eqref{merger_rate} to the minimum time delay of $20$ Myr and the maximum fixed to the Hubble time\til\cite{Meacher:2015iua}.  Finally, the normalization of the merger rate at $z=0$, {\em i.e.} $R_{\rm m,0}$, is fixed to the best-fit value obtained by the LIGO/Virgo/KAGRA collaboration:  $R_{m}(z=0)=105.5^{+190.2}_{-83.9}\ \mbox{Gpc}^{-3} \mbox{yr}^{-1}$\til\cite{KAGRA:2021duu}.

Once the redshift is extracted from the probability distribution in equation\til\eqref{rate:unit_of_redshift}, we can assign a fiducial luminosity distance, $d_L^{fid}(z)$, through the equation\til\eqref{eq: luminosty_distance general} with the $E(z)$ given in equation\til\eqref{eq: E_z LCDM}, that is, we are assuming flat $\Lambda$ CDM cosmology as the fiducial cosmological model and setting $\Omega_{\Lambda,0} =1- \Omega_{\rm m,0}$. Cosmological parameters were fixed to the best-fit values obtained by fitting the CMB power spectrum measured by the {\em Planck} satellite: $ H_0 = 67.66\ \mbox{km}\ \mbox{s}^{-1}\mbox{Mpc}^{-1},\ \Omega_{\rm m,0}= 0.3111 $\til\cite{Planck:2018vyg}.

To generate a GW catalogue, firstly, we compute the total number of observable BNS mergers, $N$, which is
\begin{equation}
    N = T_{obs}\ \mathcal{D} \int_{0}^{10} \frac{R_m (z)}{1+z}\frac{dV(z)}{dz}dz\,,
\end{equation}
where $\mathcal{D}$ is the duty cycle and $T_{obs}$ is the observation time. We set the duty cycle of the detectors to be 85\% and generate events for 1, 5, and 10 years of observations. Secondly, we assume an isotropic distribution in the sky angles $\theta$ and $\phi$, and a uniform distribution in the orientation angle $\cos i$ and the polarization $\psi$. Finally,  to generate the synthetic signal self-consistently with our choice of $R_{\rm m,0}$, we follow the LVK collaboration and set a uniform range of mass of NS in the interval $[1, 2.5]\ M_\odot$\til\cite{KAGRA:2021duu}. 

We focus on the following set of GW parameters  $\mathbf{\Theta}=\{m_1,\, m_2,\, d_L,\, \theta_{jn},\, t_c,\,\phi_c,\,\psi,\, ra,\, dec,\}$.
In particular, we have the sky angles ($ra$, $dec$), the inclination $\theta_{jn}$, the polarization angle $\psi$, the phase at coalescence $\phi_c$, the coalescence time $t_c$, and the luminosity distance of the source, $d_L$ and, finally,  the masses $m_1$ and $m_2$ of binary system.
The expected accuracy on the GW parameters is computed using the Fisher information matrix (FIM). Specifically, to compute the expected SNR and the FIM, we use the publicly available code \texttt{GWFISH}\til\cite{Dupletsa:2022scg}. Finally, the detector output, $s(t)$, is given by
\begin{equation}
    s(t) = h(t,\mathbf{\Theta}) + n(t)\,,
\end{equation}
where $h(t,\mathbf{\Theta})$ is the GW signal which depends on a set of GW parameters $\mathbf{\Theta}$, and $n(t)$ is the detector noise which is assumed to be Gaussian and stationary with zero mean. Thus, the likelihood is 
\begin{equation}
    \label{eq: GW likelihood}
    \mathcal{L}_{GW}(\mathbf{\Theta})\propto\exp{\left\{-\frac{1}{2}( s_i - h_i(\mathbf{\Theta})|s_i - h_i(\mathbf{\Theta}) \right\}} \,.
\end{equation}
Here $(\cdot |\cdot)$ is the inner product defined by
\begin{equation}\label{eq: inner prod}
    (a|b) = 4 \Re{\biggl[\int_{f_{min}}^{f_{max}} \frac{\tilde{a}^*(f) \Tilde{b}(f)}{S_{n}(f)}df\biggr]}\,,
\end{equation}
where $f_{min}$ and $f_{max}$ are the minimum and maximum frequency of the detector; $S_n$ is the detector noise power spectral density (PSD); and, finally, the tilde denotes a temporal Fourier transform.
Starting from equation \eqref{eq: inner prod}, we can define the SNR, $\rho$, as follows
\begin{equation}
    \rho = \sqrt{(h|h)}\,.
\end{equation}
The Fisher Matrix formalism allows us to write the Eq.\til\eqref{eq: GW likelihood} as follows\til\cite{Dupletsa:2024} 
\begin{equation}
    \mathcal{L}_{GW}(\mathbf{\Theta}) \propto \exp{\left\{ -\frac{1}{2}\Delta\Theta^a\mathcal{F}_{ab}\Delta\Theta^b \right\}},
\end{equation}
where $\Delta\mathbf{\Theta} = \mathbf{\Theta} - \bar{\mathbf{\Theta}}$, with $\bar{\mathbf{\Theta}}$ are the {\em true} values of GW parameters, and $\mathcal{F}_{ab}$ is the Fisher matrix defined as
\begin{equation}
    \mathcal{F}_{ab} =\left(\frac{\partial h}{\partial \Theta^a}\middle| \frac{\partial h}{\partial \Theta^b}\right)_{\mathbf{\Theta} = \bar{\mathbf{\Theta}}}\,.
\end{equation}
Once we have computed $\mathcal{F}_{ab}$, we can estimate the expected statistical uncertainties on the parameters $\mathbf{\Theta}$ as: 
\begin{equation}\label{eq:fim}
    \sigma_{\Theta^a}=\sqrt{(\mathcal{F}^{-1})^{ab}}\,.
\end{equation}
The synthetic GW signals are generated with the \texttt{TaylorF2} waveform, the simplest frequency-domain approximant\til\cite{Buonanno:2009zt}, and only GWs events with SNR$>9$ are retained. 
The choice of \texttt{TaylorF2} waveform for BNS events is supported by the comparison with respect to other waveforms made in\til\cite{Dupletsa:2022scg}.

Once we have generated the GW catalog, we extract the {\em observed} mock luminosity distance from the Gaussian distribution $\mathcal{N}(d_{L}^{fid},\sigma_{ d_{L}})$, and then use it to constrain the cosmological parameters through the redshift-luminosity distance relation.
The variance $\sigma^2_{ d_{L}}$ is given by the sum of three different noise sources:
\begin{equation}\label{sigma_dl}
 \sigma_{d_L}^2={\sigma_{inst}^2+\sigma_{lens}^2 +\sigma_{pec}^2}\,.
\end{equation}
The first one, $\sigma_{inst}$, is the contribution due to the instrument and it is computed from the $i-th$-diagonal element in the FIM in the equation \eqref{eq:fim} corresponding to the luminosity distance in the $\mathbf{\Theta}$ array. 
Then, we also consider two systematic effects $\sigma_{lens}$ and $\sigma_{pec}$. The first one is due to the weak lensing distortions and is parameterized as follows
\begin{equation}
    \sigma_{lens}=0.066\left(\frac{1-(1+z)^{-0.25}}{0.25}\right)^{1.8}d_L(z)F(z),
    \end{equation}
where $F(z)= 1- \frac{0.3}{\pi /2}\arctan{\frac{z}{z_*}}$, with $z_*=0.073$\til\cite{Speri:2020hwc}.
Finally, the second systematic effect is related to the peculiar velocities and can be approximated with the following fitting formula
\begin{equation}
\sigma_{pec}=\left[ 1+\frac{c(1+z)^2}{H(z)d_L (z)}\right]\frac{\sqrt{\langle v^2\rangle}}{c}d_L (z)\,,
\end{equation}
where the averaged peculiar velocity $\sqrt{\langle v^2\rangle}$ is set to  $500$ km/s, in agreement with the observed values in galaxy catalogs\til\cite{Cen:1998nj}. 

In our analysis, we will consider both ET and CE GW detectors. We assume that ET will be a triangular-shaped detector made up of 3 independent interferometers (ET-1, ET-2, ET-3) co-located in Italy. The PSD is described by the 10 km arm ET-D noise curve model\footnote{The most recent PSD for ET can be found at \url{https://www.et-gw.eu/index.php/etsensitivities} }. However, CE consists of two independent L-shaped detectors placed in the United States: the first in Idaho and the second in New Mexico, with arm lengths of 40 and 20 km, respectively\footnote{The most recent PSD of CE can be found at \url{https://dcc.cosmicexplorer.org/CE-T2000017/public}.}. Table \ref{tab:detectors} lists the main features of the interferometers: localization, orientation, and the lowest frequency of the power spectral density (as reported in Table III of\til\cite{Borhanian:2020ypi}). 
\begin{table}
    \setlength{\tabcolsep}{0.35em}
     \renewcommand{\arraystretch}{1.3}
    \begin{tabular}{c c c c c c }
    \hline
    Detector & Latitude &Longitude& \shortstack{x-arm\\ azimuth} & \shortstack{y-arm\\  azimuth}& $f_{ini}\, [\text{Hz}]$\\
    \hline
    ET-1 & 0.7615 &  0.1833 & 0.3392 &5.5752& 2\\
    ET-2 & 0.7629 & 0.1841 & 4.5280 & 3.4808 & 2 \\
    ET-3 & 0.7627 & 0.1819 & 2.4336& 1.3864 & 2\\
    CE-ID & 0.7649 & -1.9692 &1.5708 & 0 & 5 \\
    CE-NM &0.5787 & -1.8584 &2.3562 & 0.7854& 5 \\
    \hline
    \end{tabular}
    \caption{Localization and lowest frequency of the detectors considered \cite{Borhanian:2020ypi}. In particular, CE-ID refers to the CE located in Idaho and CE-NM to the one located in New Mexico. }
    \label{tab:detectors}
\end{table} 

In our statistical analysis, we will consider the following detector network configurations:
ET, ET + CE, ET + 2CE. In particular, in the ET+CE configuration, we consider the CE located in Idaho with 40 km arm length.
\begin{table*}
    \centering
    \setlength{\tabcolsep}{0.4em}
    \begin{tabular}{c|c|c|c||c|c|c}
    \hline
    $\theta_v$& \multicolumn{3}{c||}{Any $\theta_v$}&\multicolumn{3}{c}{$\theta_v<15^{\circ}$}\\
    \hline
        Network & ET & ET+CE  & ET+2CE & ET & ET+CE  & ET+2CE \\
        \hline
        $N_{\rm det}$ &30365& 81051 &92170 & 2458 & 4582 & 4938 \\
        $N_{\rm det}(\Delta\Omega < 1000\, \text{deg}^2)$ &789  & 76548 & 91322& 150 & 4302 & 4867 \\
        $N_{\rm det}(\Delta\Omega < 100\, \text{deg}^2)$ & 133 & 46898 &69570 & 23 &  3713 &4543 \\
        $N_{\rm det}(\Delta\Omega < 40\, \text{deg}^2)$ &  41& 19676 &35571 & 8 & 2217 & 3265\\
        $N_{\rm det}(\Delta\Omega < 10\, \text{deg}^2)$ &  8& 2720 & 6249& 1 & 404 &794 \\
        \hline
    \end{tabular}
    \caption{Number of events detected by ET, ET+CE and ET+2CE after one year of observations. We only report the GW events with sky localization better than 10, 40, 100 and 1000 deg$^2$. Moreover, we distinguish BNS mergers with a generic orientation or a viewing angle $\theta_v < 15^{\circ}$. For the latter class of events, we have a more significant probability of detecting a coincident GRB.}
    \label{tab: detected event}
\end{table*}
In Table \ref{tab: detected event}, we report the number of detected events after one year of observations for the different network configurations. Furthermore, we list the expected number of detections for four sky-localization uncertainty thresholds, namely 10, 40, 100, and 1000 deg$^2$ at the 90\% confidence level. Finally, we give the number of GW events with a viewing angle, $\theta_v$, smaller than $15^{\circ}$\footnote{The viewing angle $\theta_v$ is the $\min(i, 180^{\circ} - i)$.}. Only a fraction of those events are expected to produce detectable high-energy emissions powered by the GRB relativistic jet, which is assumed to be perpendicular to the orbital plane.

\subsection{Electromagnetic counterparts}\label{subsec: EM cunterpart}

Following~\cite{Ronchini:2022gwk,Branchesi:2023mws}, we want to analyze the multi-messenger capabilities of the 3G detectors when synergies with electromagnetic (EM) observatories are considered. In particular, we will focus on the $\gamma$-rays produced by the prompt emission of a GRB; the X-rays due to the afterglow of GRB; and, the optical emission produced by a KN.

\subsubsection{Prompt Emission}\label{subsec: prompt emission}

Let us start by assuming that only 20\% of the BNS mergers produce a detectable jet~\cite{Ronchini:2022gwk,Branchesi:2023mws}. We record a joint detection if the electromagnetic counterpart is evaluated under the assumption of a universal jet structure, as it was for GRB 170817A, which means we assume the same jet features for all GRBs\til\cite{Postnov:1999gj,Salafia:2022dkz}. The angular structure of the local emissivity, $\epsilon(\theta_v)$, and the bulk Lorentz factor profile, $\Gamma(\theta_v)$, are parametrized in terms of the viewing angle $\theta_v$ as follows:
\begin{align}
\label{eq:epsilon} \epsilon(\theta_v) &= \frac{4\pi \epsilon_c}{1+\left( \frac{\theta_v}{\theta_c}\right)^{s_{\epsilon}}},\\
\label{eq:Gamma} \Gamma(\theta_v) &= 1 + \frac{\Gamma_0 - 1 }{1+\left( \frac{\theta_v}{\theta_c}\right)^{s_{\Gamma}}}.
\end{align}
Following~\cite{Ronchini:2022gwk}, we set $s_{\epsilon} = s_{\Gamma} = 4$, $\theta_c = 3.4^{\circ}$, $\Gamma_0 = 500$, and $4\pi \epsilon_c = 3\times 10^{53}\ \text{erg}$.

To classify the detectability of an event, we need to estimate the flux related to the GRB prompt emission. Thus, the photon flux received by the detector is
\begin{equation}\label{eq:flux}
\mathcal{F} = \frac{L_{\rm }}{4\pi d_L^2} \times k(z),
\end{equation}
where the isotropic equivalent luminosity $L_{\rm }$ is related to the isotropic equivalent energy $E_{\rm }$ through the relation:
\begin{equation}\label{eq:L_iso}
L_{\rm } = \frac{2 E_{\rm }}{\langle t_{GRB}\rangle},
\end{equation}
with $\langle t_{GRB}\rangle = 22\ \text{s}$ \cite{Salafia:2015vla}. In the equation~\eqref{eq:flux}, we also consider the $k$-correction due to the redshifted photon energy when travelling from the source to the detector:
\begin{equation}
k = \frac{\int_{E_1(1+z)}^{E_2(1+z)}N(E)dE}{\int_{E_1}^{E_2}N(E)dE}.
\end{equation}
Here, $E_1$ and $E_2$ represent the detector’s energy window, and $N(E)$ is the observed GRB photon spectrum modelled by the band function\til\cite{Band:1993eg,Nava:2010ig}. Then, we extract the peak energy $E_p$ from a log-normal distribution \cite{Ronchini:2022gwk}:
\begin{equation}
P(\log(E_p)) \propto \exp\left( -\frac{(\log(E_p)-\log(\mu_E))^2}{\sigma_E^2}\right),
\end{equation}
with $\log_{10}(\mu_E/\text{keV})=3.2$ and $\sigma_E=0.36$. Finally, $E_{\rm }$ is written as \cite{Salafia:2015vla,Salafia:2019off}:
\begin{equation}\label{eq:E_iso}
E_{\rm }(\theta_v) = \int \frac{\delta^3(\theta,\phi,\theta_v)}{\Gamma(\theta)}\epsilon(\theta)d\Omega.
\end{equation}

For an off-axis observer, the Doppler factor
$\zeta$ must consider the angle between the line of sight and the velocity of each point on the surface, $\beta(\theta)$. The Doppler factor is then given by:
\begin{equation}
\zeta(\theta,\phi,\theta_v)=\frac{1}{\Gamma(\theta)[1 - \beta(\theta)\cos(\alpha(\theta,\phi,\theta_v))]},
\end{equation}
where $\beta (\theta) = \left(1- \frac{1}{\Gamma^2(\theta)}   \right)^{1/2}$ and $\cos(\alpha)= \cos\theta \cos\theta_v +\sin\theta\sin\phi\sin\theta_v$.

To evaluate the flux for the BNS in the catalogue, we estimate the $E_{\rm }$ and, consequently, the luminosity $L_{\rm }$ by using equation~\eqref{eq:E_iso} and equation~\eqref{eq:L_iso}, respectively. Finally, we obtain the flux from the equation~\eqref{eq:flux}. However, we still need to estimate the number of combined events. To do so, we consider the X/Gamma-ray Imaging Spectrometer (XGIS) on board the THESEUS satellite as a detector of the prompt emission of the GRBs produced by the BNS population in our catalog. We set the duty cycle of THESEUS-XGIS at 85\%, and the field of view (FOV) to $\sim 2$ sr in the energy band $2-150$ keV \cite{THESEUS:2017wvz}. The probability of detection is given by the ratio $\text{FOV}/4\pi \sim 0.16$, while the flux threshold for XGIS in the energy band $2-150$ keV is $3\times10^{-8}\,\text{erg}\,\text{cm}^{-2}\,\text{s}^{-1}$ \cite{Amati2021}. Finally, 
in Table~\ref{tab:number-events-combined}, we report the number of combined events obtained with the detection of the prompt emission of the GRB for the three network configurations considered. Our estimates
are comparable with the results presented in\til\cite{Ronchini:2022gwk,Branchesi:2023mws} demonstrating the effectiveness of the procedure.

\subsubsection{Afterglow}\label{subsec: afterglow}

The light curves of the afterglow emission are computed using the public Python package \texttt{afterglowpy} \cite{Ryan:2019fhz}. These light curves depend on the jet structure introduced in equations~\eqref{eq:epsilon} and \eqref{eq:Gamma}, as well as certain microphysical parameters. Specifically, these parameters include the density of the interstellar medium (ISM), $n_0$, the slope of the electron energy distribution, $p$, and the fractions of energy carried by electrons and the magnetic field ,$\epsilon_e$ and $\epsilon_B$, respectively.

Furthermore, we consider the fraction of kinetic energy transformed into radiation, denoted by $\eta$, and the angular extension of the jet wing, $\theta_w$. Following \cite{Ronchini:2022gwk}, we set the parameters to the following values: $p=2.2$, $\epsilon_e=0.1$, and $\theta_w = 15^{\circ}$; and we uniformly sample from the confidence intervals derived in\til\cite{Fong:2015oha}: $\epsilon_B \in [0.01, 0.1]$, $\eta \in [0.01, 0.1]$, and $n_0 \in [3\times 10^{-3}, 15\times 10^{-3}] \ \text{cm}^{-3}$.

We particularize our analysis for the Soft X-ray Imager (SXI) onboard the THESEUS mission to detect the X-ray emission associated with the afterglow. The SXI is a wide-field instrument capable of covering 0.5 steradians in a single observation, providing a localization accuracy of 1-2 arcmin. 
The flux threshold for SXI in the energy band $0.3-5$ keV is $1.8\times10^{-11}\,\text{erg}\,\text{cm}^{-2}\,\text{s}^{-1}$ \cite{Amati2021}. Additionally, the XGIS exhibits good sensitivity down to 2 keV and covers a larger field of view (2 steradians). Considering the complementary strengths of both instruments, we explore the combined use of SXI and XGIS. In Table~\ref{tab:number-events-combined}, we report the number of X-ray detections associated with the GW events after one, five, and ten years of observations. It should be noted that these estimates are slightly lower than the results shown in \cite{Ronchini:2022gwk,Branchesi:2023mws}, because the Python package \texttt{afterglowpy} omits the contribution of high-latitude emission\til\cite{Ascenzi:2020kxz}.

\begin{table}
    \centering
    \begin{tabular}{c|c|c|c|c}
    \hline
     \multicolumn{5}{c}{\bf Prompt Emission}\\
            \hline
        Instrument&Years & ET & ET+CE & ET+2CE\\
        \hline
        \multirow{3}{*}{\makecell{THESEUS-XGIS\\($\gamma$-ray)}} &1&13&30  &35 \\
       
        &5& 47 & 153 &171 \\
        
        &10&101  &292& 333\\
        \hline
    \hline 
    \multicolumn{5}{c}{\bf Afterglow}\\
    \hline
    Instrument & Years & ET & ET+CE & ET+2CE\\
    \hline
    \multirow{3}{*}{\makecell{THESEUS-SXI\\(X-ray)}} &1&3&4  &6 \\
    &5& 14 & 24 & 26 \\
    &10&20  &58& 67\\
    \hline
    \multirow{3}{*}{\makecell{THESEUS\\SXI+XGIS\\ (X-ray)}} &1&12&13  &16 \\
    &5& 54 & 83 &87 \\
    &10&99  &139& 148\\
    \hline
    \hline
      \multicolumn{5}{c}{\bf Kilonovae}\\
    \hline
       Instrument & Years & ET & ET+CE & ET+2CE\\
        \hline
        \multirow{3}{*}{\makecell{VRO\\(Optical)}} &1&19&802  &965 \\
        
        &5& 91 & 4192 &4952 \\
        &10&186  &8220& 9794\\
        \hline
    \end{tabular}
    \caption{We report the number of joint GW + EM detections for different network configurations after one, five, and ten years of observations. The  $\gamma$-ray signals are produced by the GRBs prompt emission detected by THESEUS-XGIS. Whereas, the X-ray emission comes from the afterglow of the GRBs detected by THESEUS-SXI alone and in combination with THESEUS-XGIS. Finally, we consider VRO as an optical detector of the KNs for which we restrict ourselves to GW events with a sky-localization uncertainty smaller than $40$ deg$^2$. }
    \label{tab:number-events-combined}
\end{table}

\subsubsection{Kilonovae Emission}\label{subsec: KN}
Finally, we also study the possibility of a joint detection of GWs and KN. 
We consider the VRO as the optical detector of the KNs associated with the GW events. The VRO is an 8.4-meter telescope featuring a wide field (9.6 degrees$^2$) camera equipped with more than three billion pixels in solid-state detectors\til\cite{LSST:2008ijt}. Its streamlined design allows for quick and precise maneuvering across expansive sections of the sky. This telescope is optimal for GWs sky localizations, providing exceptionally high-quality, and in-depth observations.

We use the \texttt{redback} pipeline to simulate the number of joint optical/GW detection\til\cite{sarin2023redback}.
In particular, we model the KN emission following the analytical KN model first introduced in\til\cite{Metzger:2016pju} and implemented in MOSFiT\til\cite{Villar:2017oya}.  It should be stressed that the KN emission is isotropic, so it can be observed regardless of the binary inclination. Following\til\cite{Branchesi:2023mws}, we select only the BNS events with a sky-localization uncertainty smaller than $40$ deg$^2$. In  Table\til\ref{tab: detected event}, we also report the number of GW events with a detectable KN counterpart in one years of observations.  We adopt the Target of Opportunity strategy described in\til\cite{Andreoni:2021epw,Cowperthwaite:2018gmx,LSST:2018bbx}. In particular, following the {\it minimal} strategy of\til\cite{Andreoni:2021epw},  we consider two observational bands ($g$ and $z$) throughout the first and second night after the merger. Then, following\til\cite{LSSTTransient:2018dvm}, we also check that we obtain comparable results considering the pair of filters $g$ and $I$.

We record a joint GW+KN event if the associated KN emission is detected at $5\sigma$ magnitude limits along 180 seconds of exposures, {\em, i.e.} $m^{lim}_{g}\sim 26$ mag and $m^{lim}_{z} \sim 24.4$ mag, during the first and second nights in both filters. 
Observations made in at least two filters play a crucial role in minimizing the impact of contaminating transients, and facilitating the identification of the KN linked to the GW signal, particularly through the analysis of colour evolution. Although spectroscopy is commonly employed in real observations to characterize sources and resolve issues related to contaminating transients, our analysis does not incorporate it.
With this strategy, it is expected to spend $\sim 3$ hours on each BNS merger. Assuming that the VRO survey will have a total of 3600 hours available per year, one might detect $\sim 1200$ KN per year\til\cite{Alfradique:2022tox}. In Table\til\ref{tab:number-events-combined}, we report the number of combined GW + KN events for one, five, and ten years of observations.

\section{Cosmological Inference}\label{sec:results}

To estimate the precision down to which 3G detectors will constrain cosmological parameters, we carried out a statistical analysis using a Markov Chain Monte Carlo (MCMC) algorithm. The underlying cosmological model in our mock data is $\Lambda$ CDM model in\til\eqref{eq: E_z LCDM}, we will refer to it as \emph{fiducial} model. Consequently, in Sect.~\ref{sec:models}, we gave the values of the cosmological parameters of the different interacting scenarios to recover the \emph{fiducial} model.

The MCMC is implemented using the \texttt{emcee} package\til\cite{emcee}, which uses the likelihood of all GW events, defined as the product of the likelihood of a single event $p(\mathbf{d}|\bm{\lambda}) = \prod_{i=1}^{N} p(d_i|\bm{\lambda})$. Here, $\bm{\lambda}$ represents the cosmological parameters of interest. Specifically, $\bm{\lambda}=\{H_0,\Omega_{\rm m,0}\}$ for the $\Lambda$CDM model, while $\bm{\lambda}=\{H_0,\Omega_{\rm m,0},\omega,\delta \ \text{or}\ \eta\}$ for the three interacting models. 
In particular, for case II and case III, the $\Omega_{\rm m,0}$ parameter is split into $\Omega_{b,0}$ and $\Omega_{\rm dm,0}$. Furthermore, $\mathbf{d} \equiv \lbrace d_i \rbrace_{i=1}^{N}$ is the mock dataset, with $N$ equal to the number of observations. Thus, the single-event likelihood is given by
\begin{equation}
    p(d_i|\bm{\lambda}) \propto \exp{\left(-\frac{1}{2}\frac{(d_i - d_L(z,\bm{\lambda}))^2}{\sigma^2_{d_L}}\right)},
\end{equation}
where $d_L(z,\bm{\lambda})$ is computed using equation\til\eqref{eq: luminosty_distance general}, and the function $E(z)$ is given in equation\til\eqref{eq: E_z LCDM} for the $\Lambda$CDM and in equations\til\eqref{eq: E(z) int DE}--\eqref{eq:E(z) int DM}--\eqref{eq:E(z) int total} for the three interacting scenarios corresponding to case I, II and III, respectively.

\begin{figure}
    \centering
   {\includegraphics[width=\columnwidth]{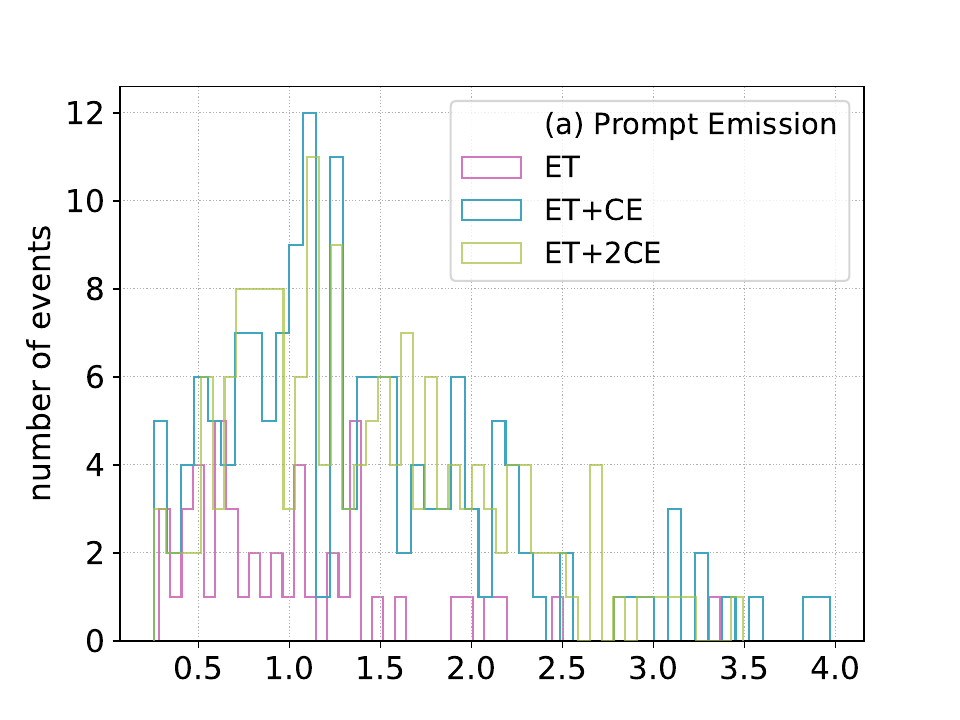}}
     {\includegraphics[width=\columnwidth]{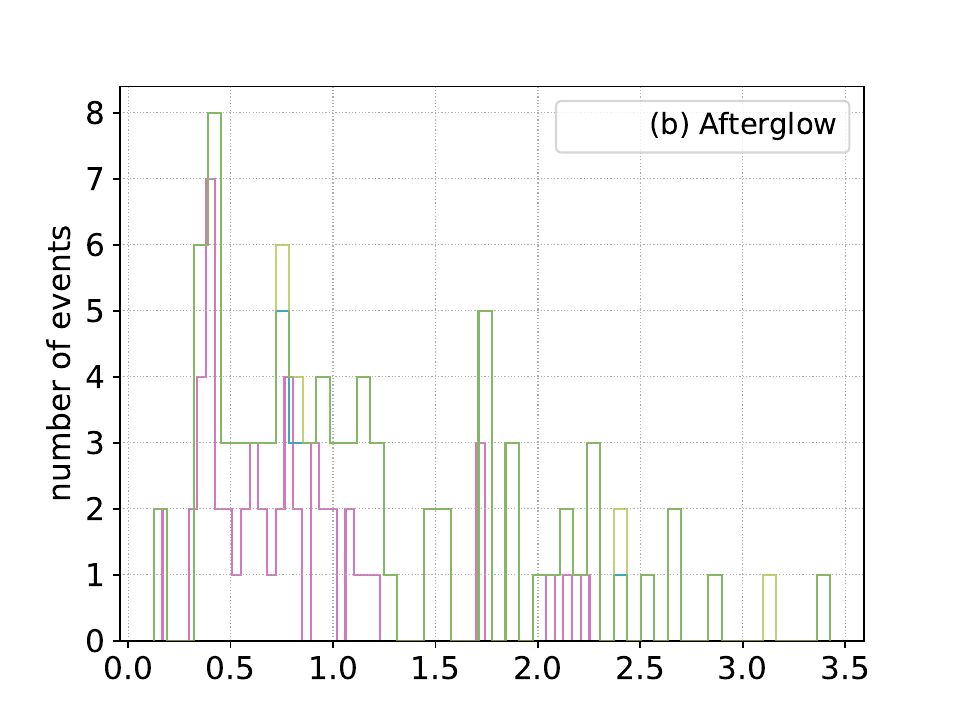}}
     {\includegraphics[width=\columnwidth]{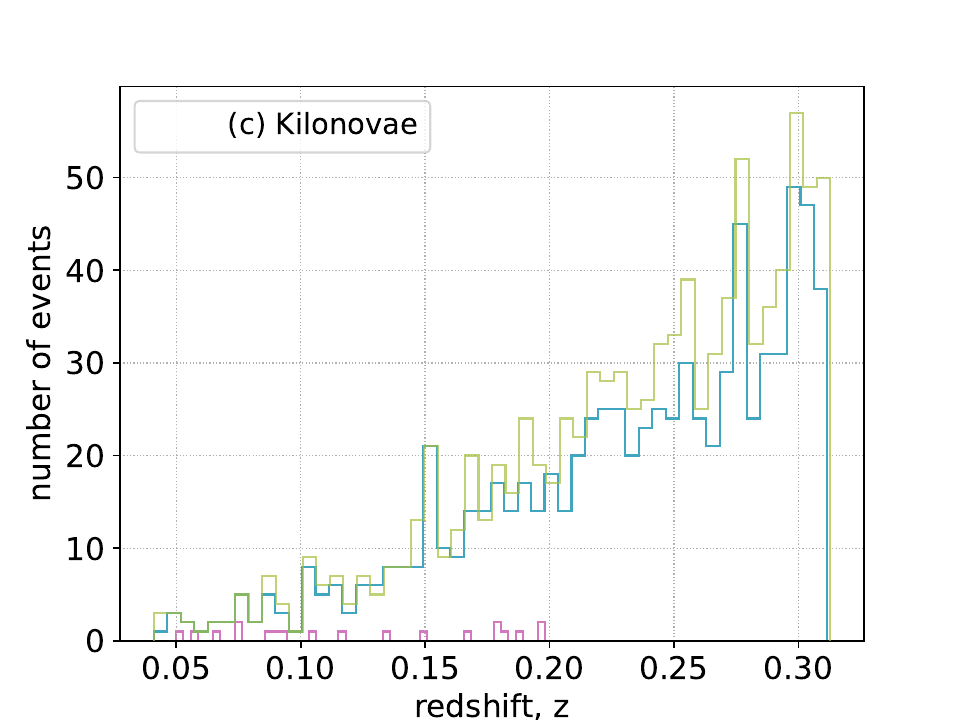}}
    \caption{Figure depicts the redshift distribution for the joint GW+EM events. In particular, we consider in panel {\em (a)} the prompt emissions detected by THESEUS-XGIS; in panel {\em (b)} the afterglows detected by THESEUS-XGIS +SXI; and in panel {\em (c)}  the kilonovae detected by VRO. We consider five years of observations for the events detected jointly with THESEUS while, for the GW+KN catalogue, we use one year of observation.
    In each panel, the histograms from the different network configurations are depicted in magenta, azure, and green for the ET, ET+CE and ET+2CE networks, respectively. }
    \label{fig: event distribution}
\end{figure}

Applying Bayes' theorem, we can write the posterior distribution, $p(\bm{\lambda}|\mathbf{d})$, of the cosmological parameters in terms of the likelihood function and the prior distribution $\pi (\bm{\lambda})$ as follows:
\begin{equation}
    p(\bm{\lambda}|\mathbf{d}) \propto p(\mathbf{d}|\bm{\lambda})\pi (\bm{\lambda}).
\end{equation}
In our analysis, we set the following priors on cosmological parameters: $\mathcal{U}(45,90)$ for $H_0$, $\mathcal{U}(0,1)$ for $\Omega_{\rm m,0}$,   $\mathcal{U}(-5,0)$ for w, $\mathcal{U}(-5,5)$ for $\delta$ and $\mathcal{U}(-5,5)$ for $\eta$.
We run the MCMC algorithm considering 5 years of observations for the joint detection of GW events with a detected GRB, facilitated by THESEUS-XGIS for the PE and THESEUS-SXI+XGIS for the AF, and one year of observation for the GW events with the associated KN detected by VRO. Additionally, we report the results for the three detector network configurations: ET, ET+CE, and ET+2CE. In Fig.\til\ref{fig: event distribution}, we depict the event redshift distribution for the different observational scenarios considered.

\begin{table}[]
    \centering
    \begin{tabular} { c |c| c c c}
    \hline
 {\bf Parameter} &  {\bf Network} & {\bf PE} &  {  \bf AF} & {\bf KN}\\
\hline 
&ET &  $65.1^{+3.5}_{-3.1}        $ & $65.3^{+3.4}_{-3.4}              $ & $67.2^{+1.6}_{-1.1}        $\\
{\boldmath$H_0$} & ET+ CE& $68.4^{+1.7}_{-1.7}              $ & $64.8^{+2.7}_{-2.4}        $ & $67.60^{+0.17}_{-0.17}            $\\

 & ET+2CE& $68.2^{+1.1}_{-1.1}               $ & $66.7^{+2.4}_{-2.4}$ & $67.63^{+0.14}_{-0.14}             $\\

\hline
&ET& $0.43^{+0.11}_{-0.19}      $ & $0.40^{+0.09}_{-0.15}    $ & $0.39^{+0.14}_{-0.35}      $\\
{\boldmath$\Omega_{\rm m,0}   $} &ET+ CE& $0.31^{+0.05}_{-0.06}   $ & $0.43^{+0.07}_{-0.11}    $ & $0.32^{+0.02}_{-0.02}        $\\
& ET+2CE & $0.31^{+0.03}_{-0.04}   $ & $0.34^{+0.06}_{-0.09}  $ & $0.31^{+0.02}_{-0.02}     $\\
\hline
\end{tabular}
\caption{The median value and the 68\% confidence level of the posterior distributions of the parameters of the $\Lambda$CDM model for each detector network configuration and for the PE, AF and KN joint detection.}\label{tab: result LCDM}
\end{table}

\section{Results}

\begin{figure*}
    \centering
    \subfigure[Prompt Emission]{\includegraphics[width=0.3\textwidth]{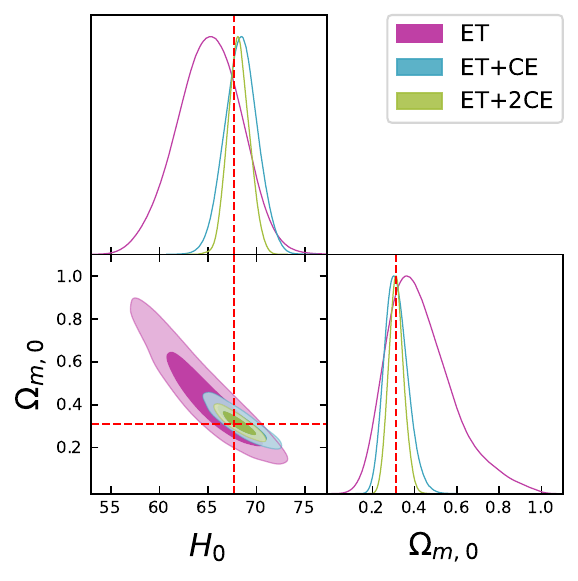}}
         \subfigure[Afterglow]{\includegraphics[width=0.3\textwidth]{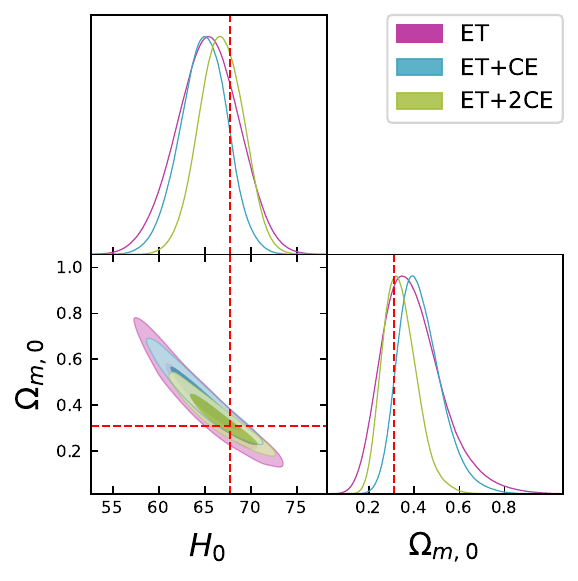}}
             \subfigure[Kilonovae]{\includegraphics[width=0.3\textwidth]{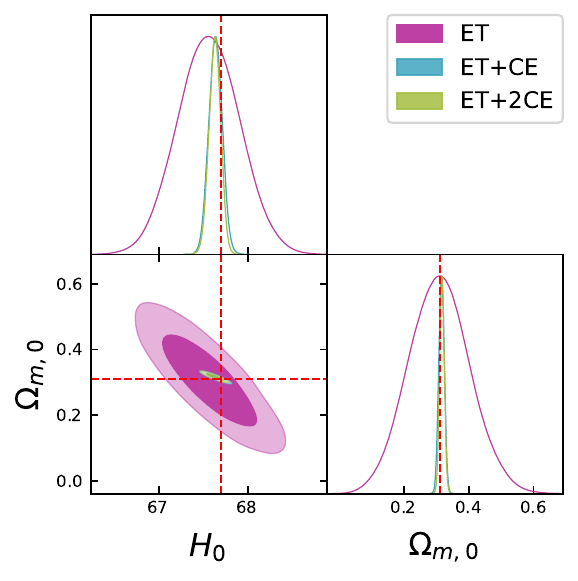}}
    \caption{Figure illustrates the 68\% and 95\% of confidence level obtained from the posterior distribution of the cosmological parameters of the $\Lambda$CDM model for the joint detection of GW and EM events.
   As an EM counterpart, we take into account (a) the prompt emissions detected by THESEUS-XGIS, (b) the afterglows detected by THESEUS-XGIS +SXI and (c) the KN detected by VRO.
In each contour plot, we represent with violet, azure, and green distributions and filled areas the results of the different network configurations: ET, ET+CE, ET+2CE, respectively. The different levels of transparency of the contours correspond to the 68\% and 95\% confidence levels with the darkest and lightest colors, respectively. Finally, the vertical dashed red line indicates the value of the {\em fiducial} cosmological parameters.}
    \label{fig: result LCDM}
\end{figure*}

\begin{figure*}
\textbf{CASE I}\\ 
    \centering
    {\includegraphics[width=0.32\textwidth]{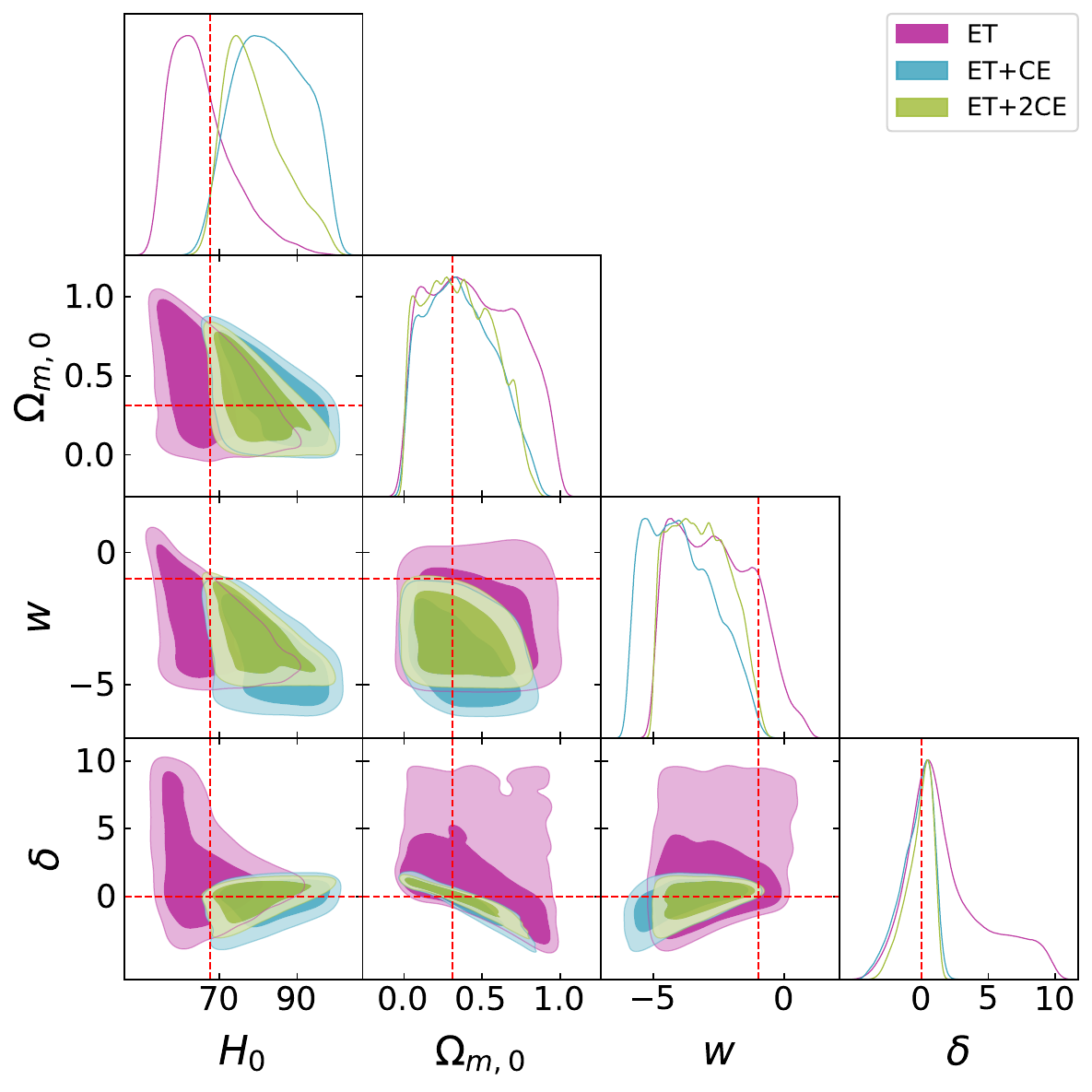}}
    {\includegraphics[width=0.32\textwidth]{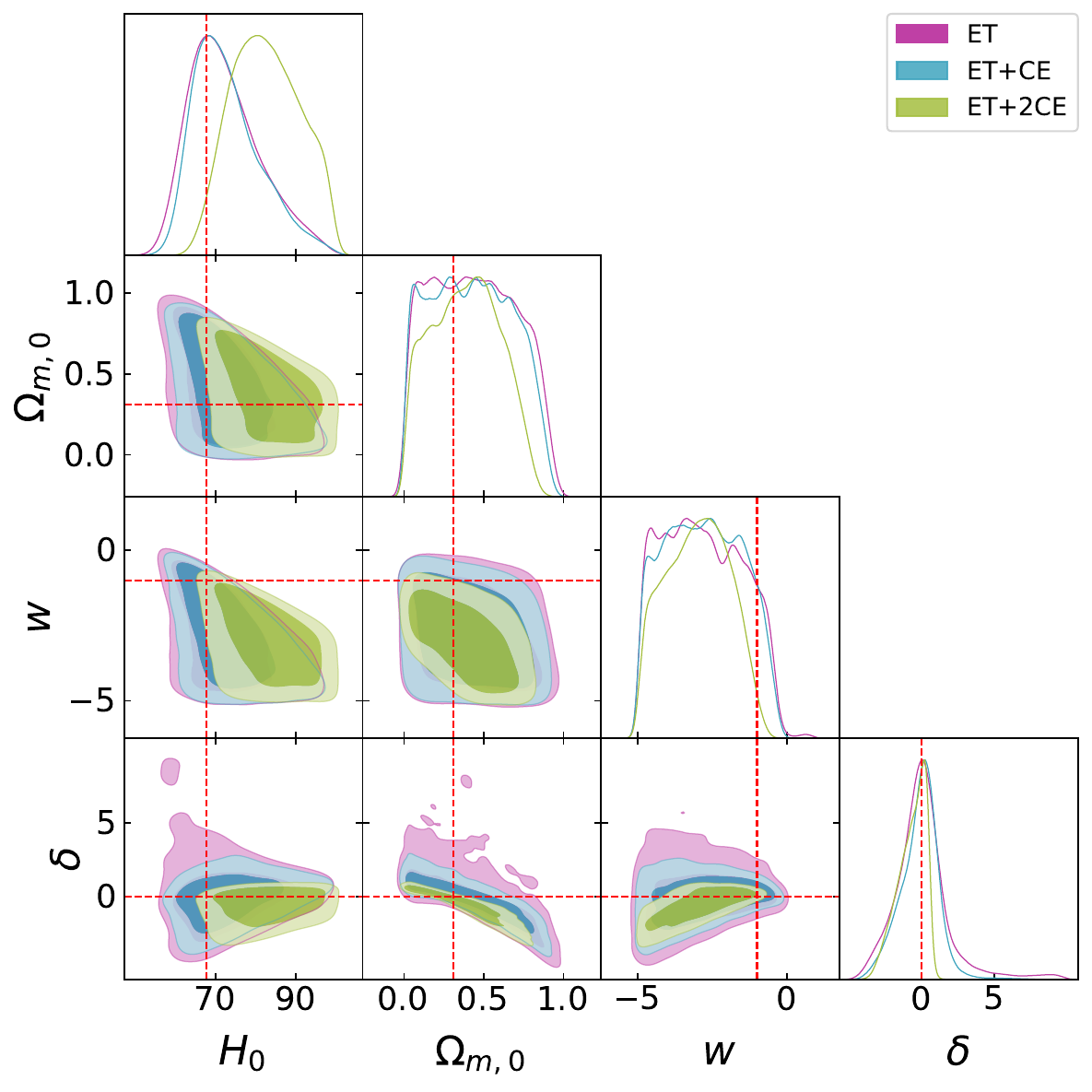}}
    {\includegraphics[width=0.32\textwidth]{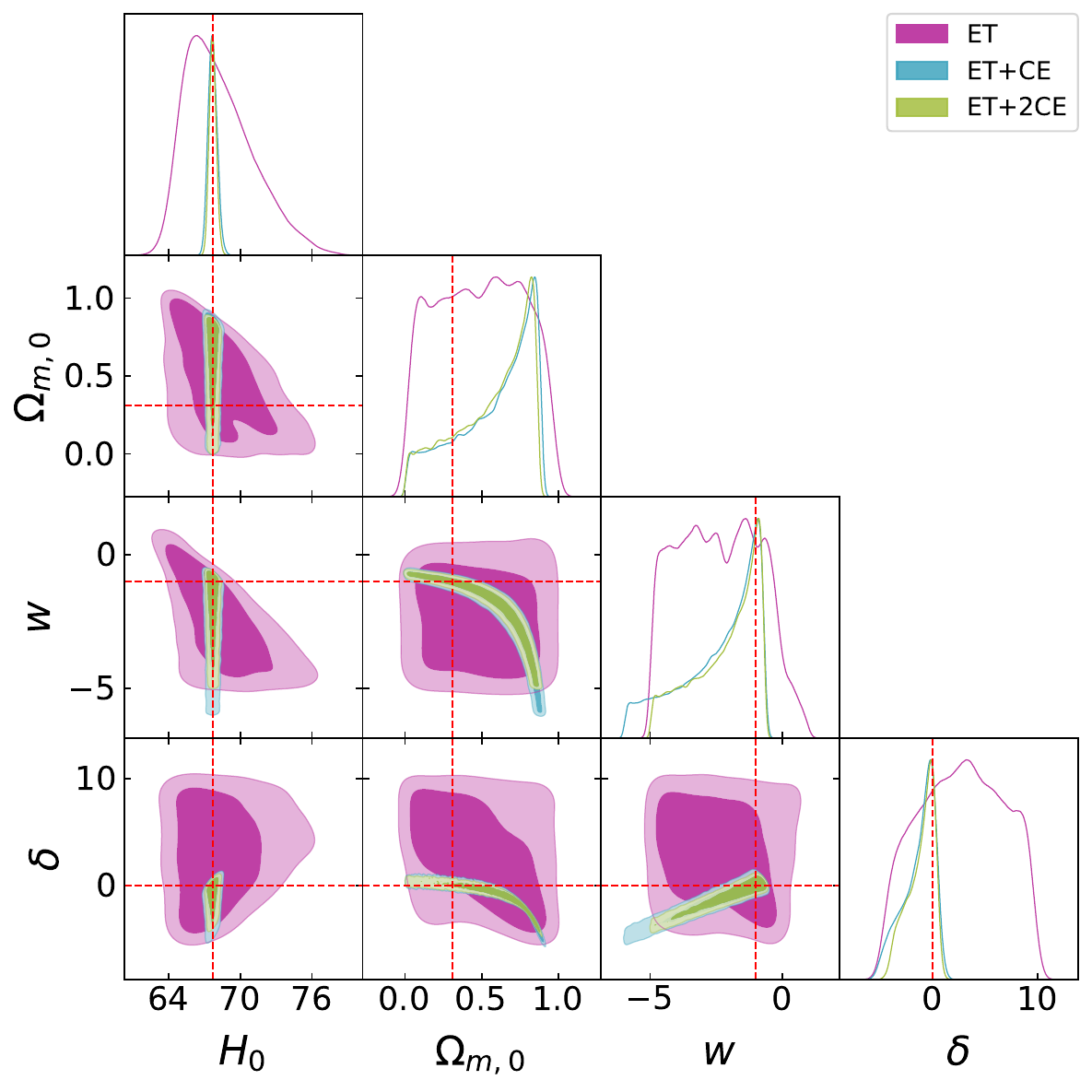}}\\
\textbf{CASE II}\\ 
    {\includegraphics[width=0.32\textwidth]{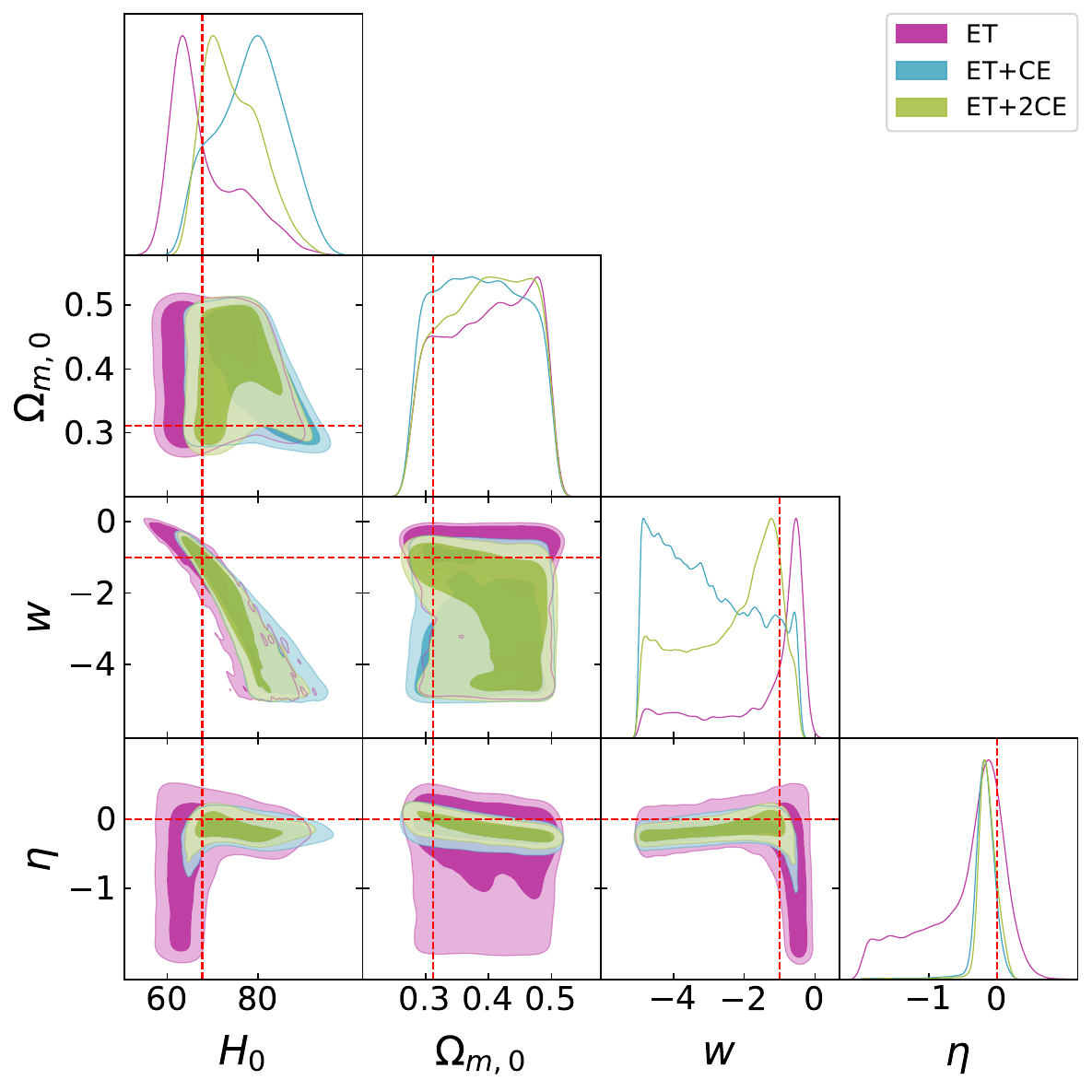}}
    {\includegraphics[width=0.32\textwidth]{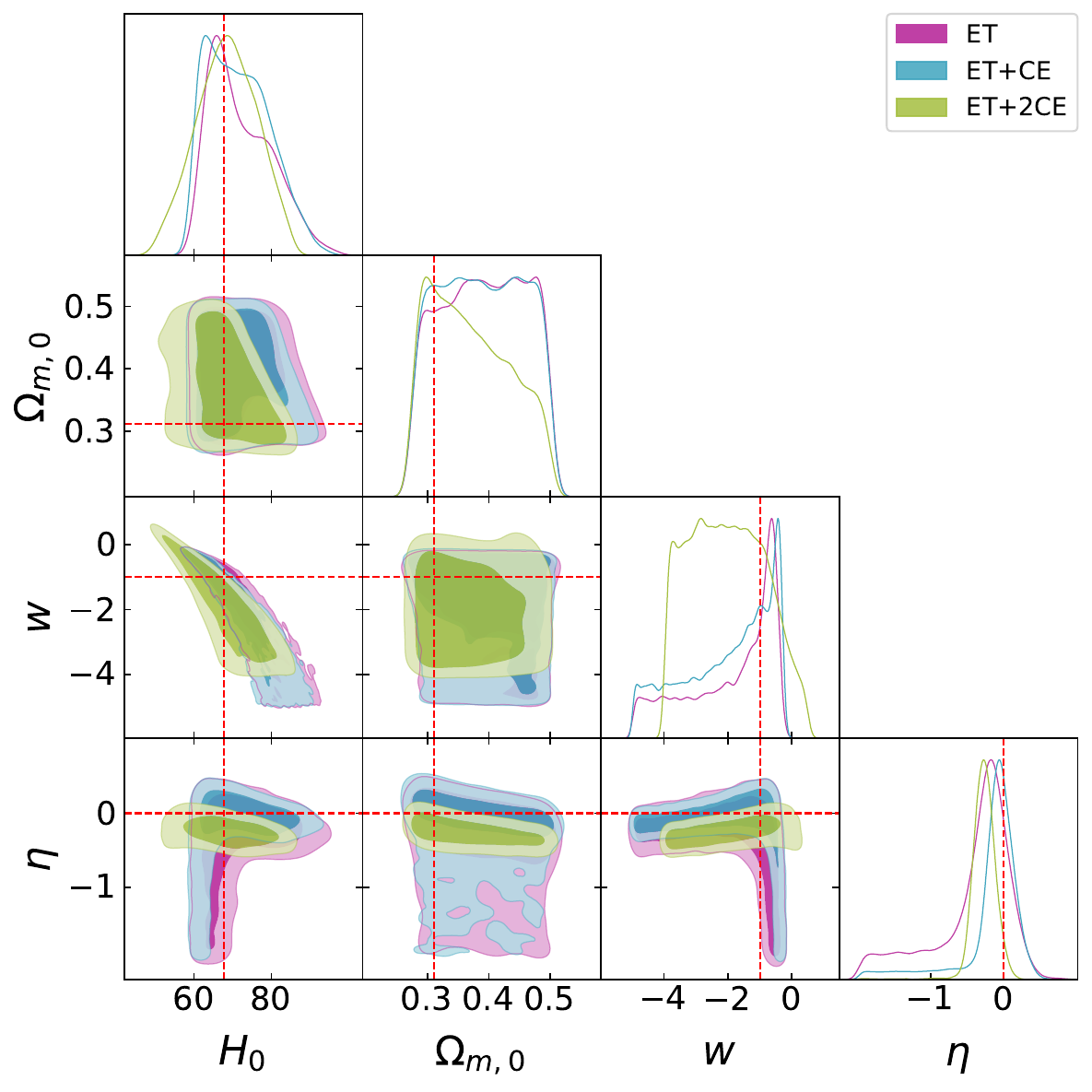}}
    {\includegraphics[width=0.32\textwidth]{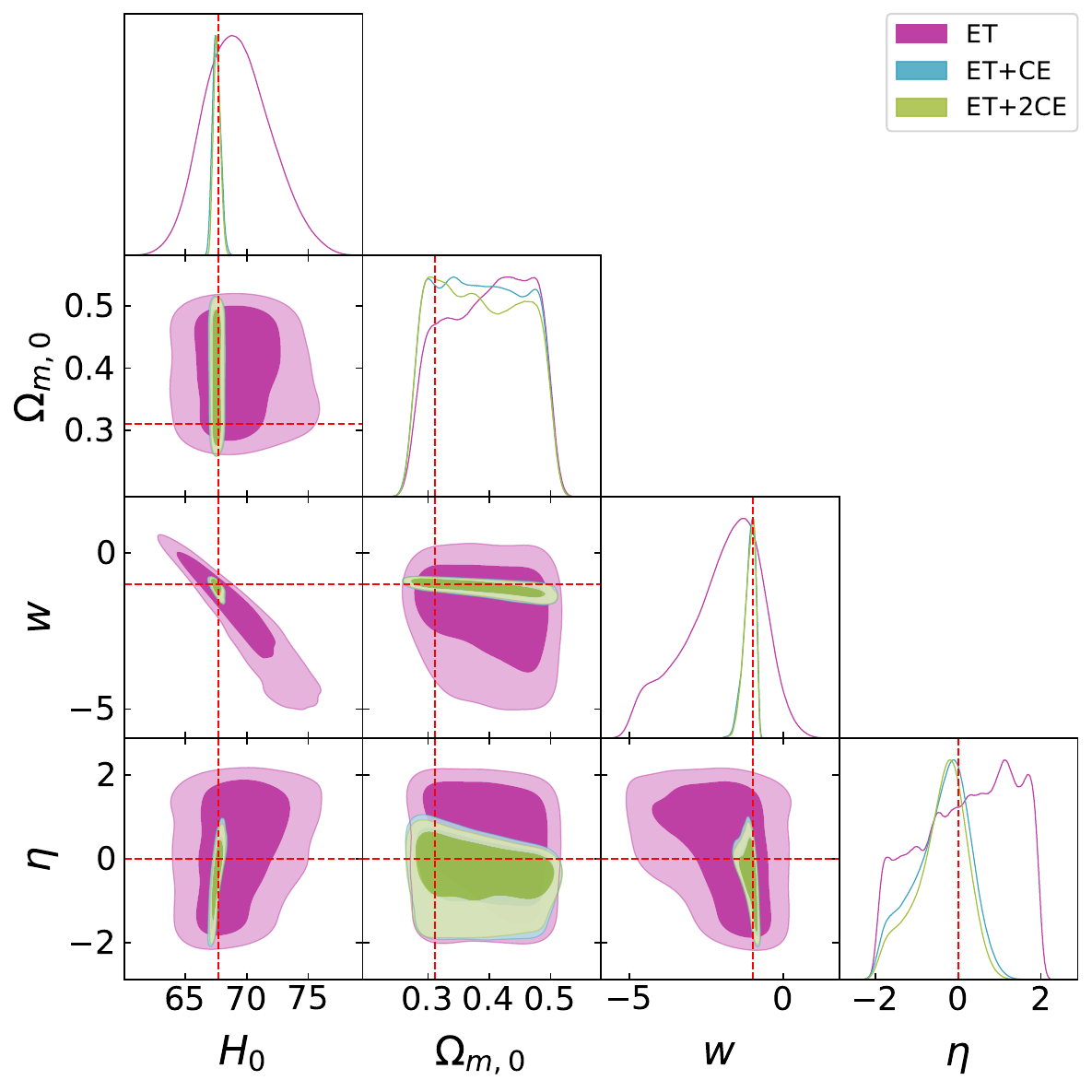}}\\
\textbf{CASE III}\\ 
    \subfigure[Prompt Emission]{\includegraphics[width=0.32\textwidth]{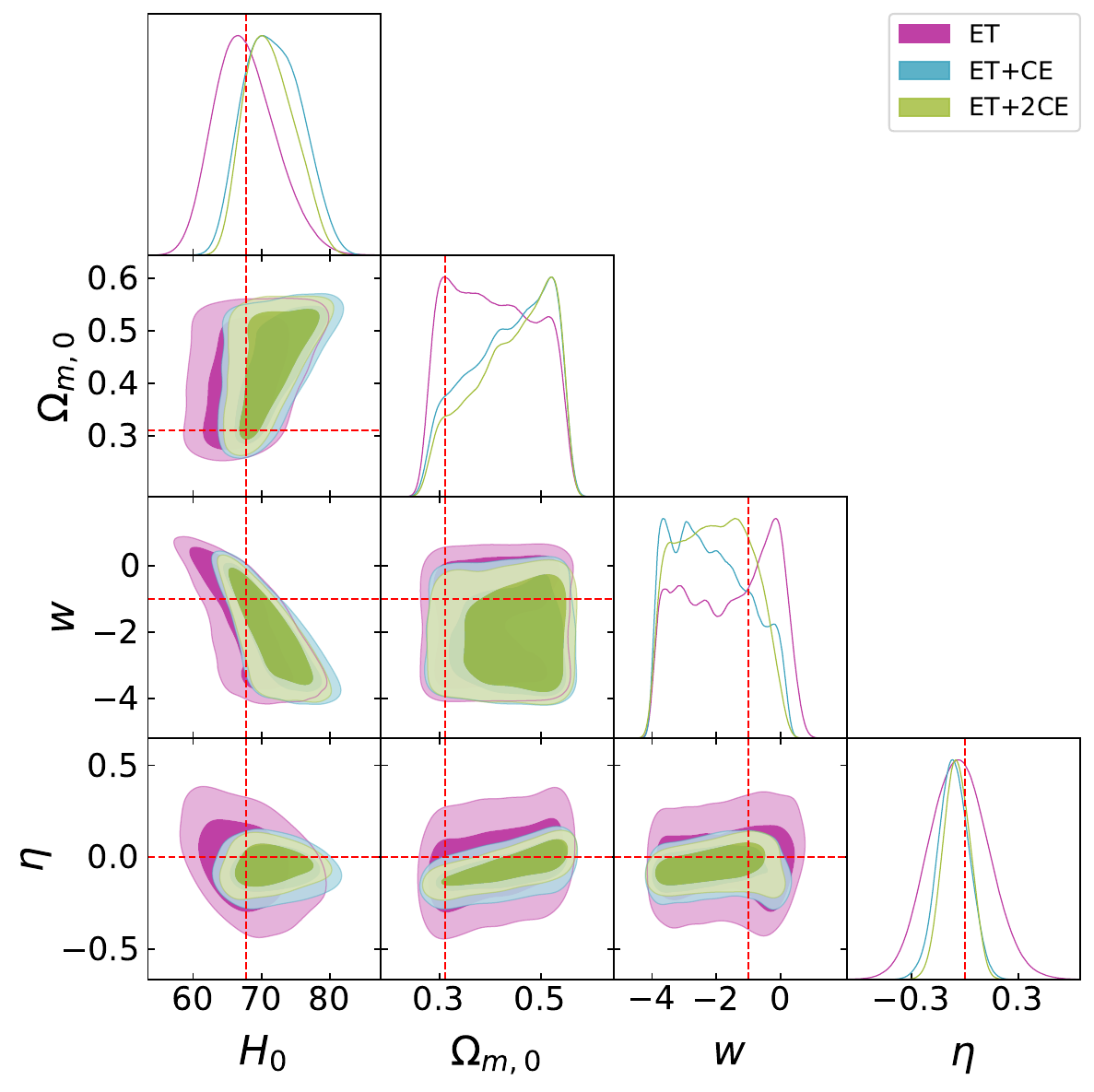}}
         \subfigure[Afterglow]{\includegraphics[width=0.32\textwidth]{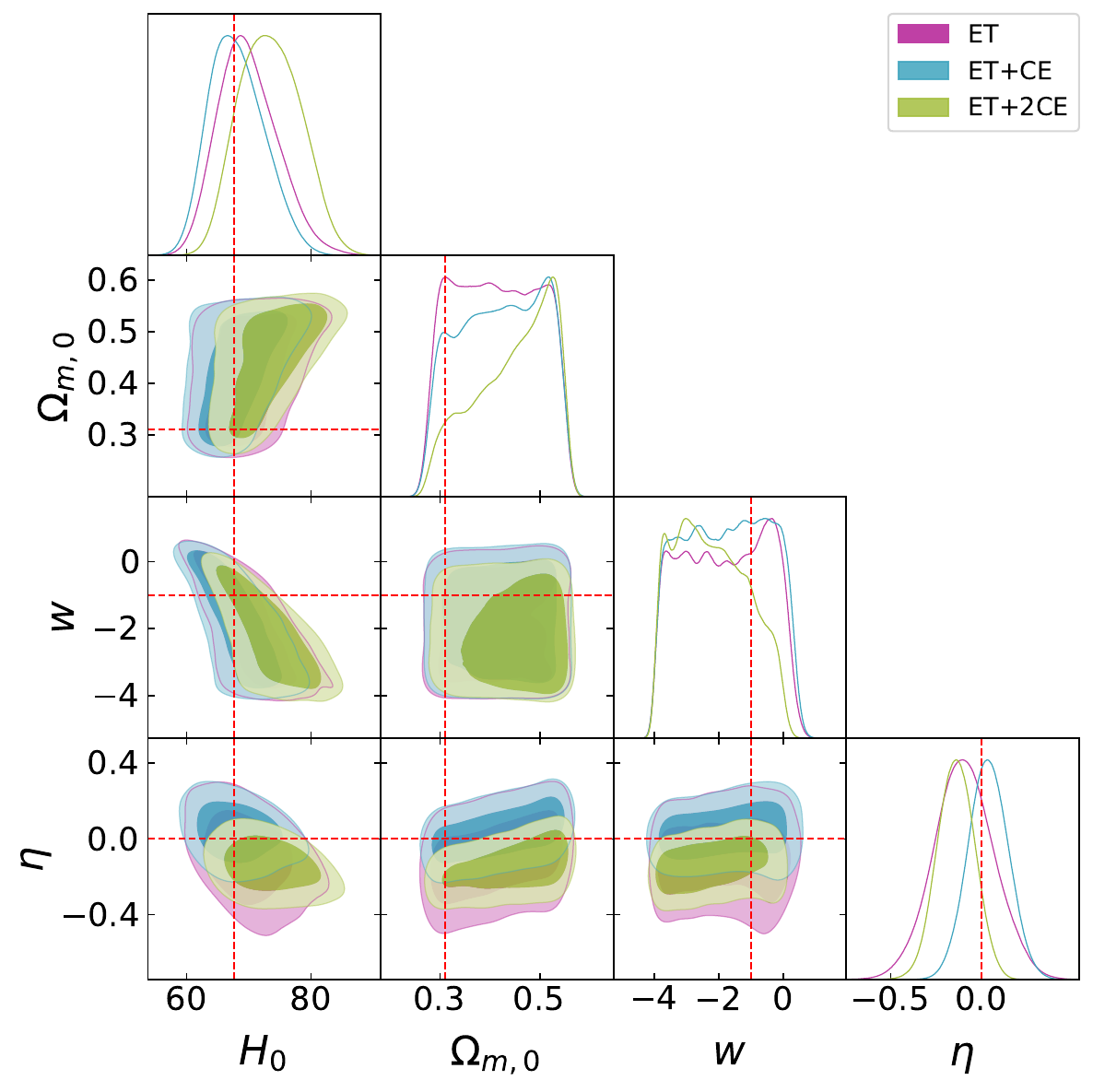}}
             \subfigure[Kilonovae]{\includegraphics[width=0.32\textwidth]{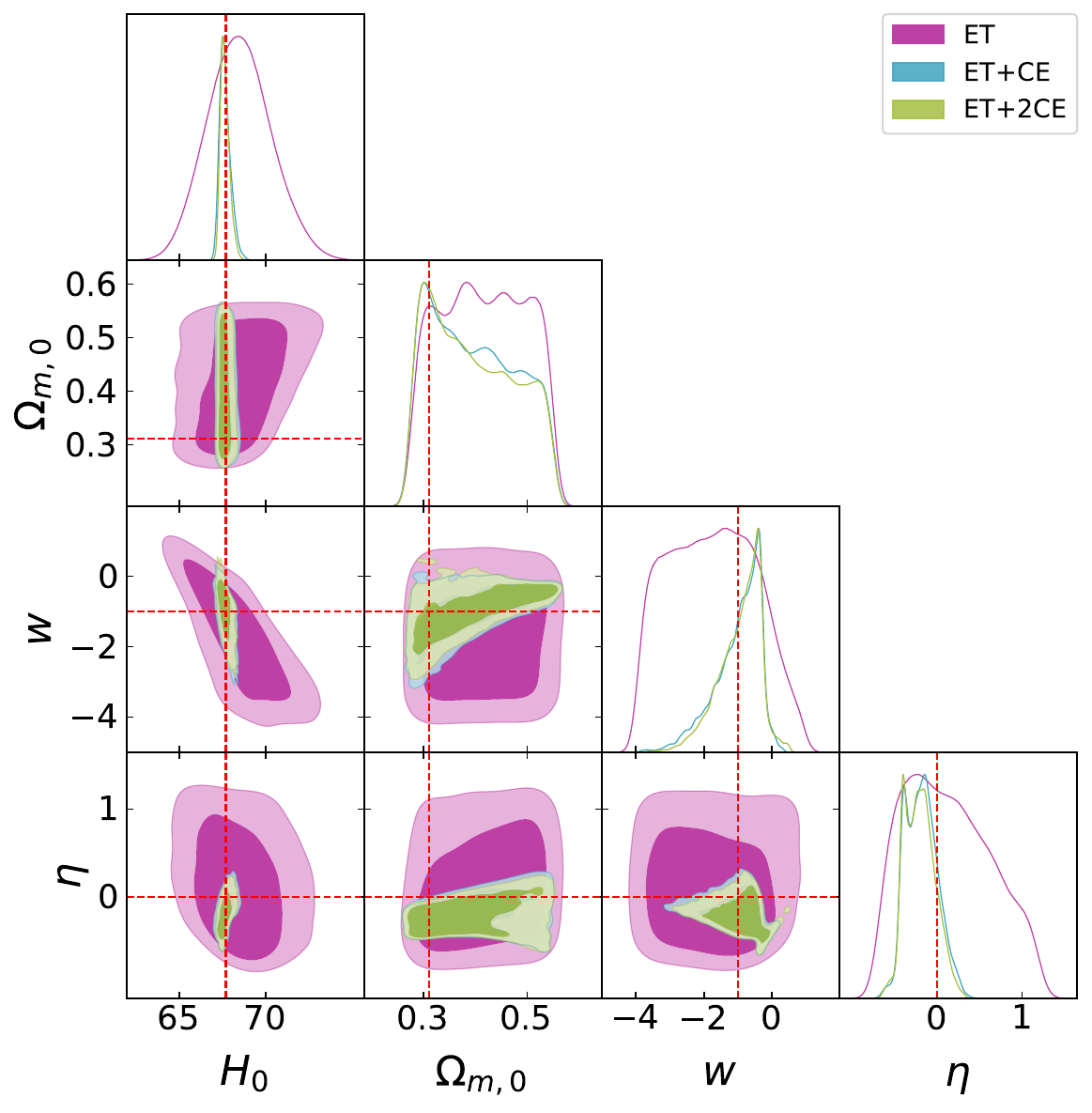}}\\
    \caption{The same of Fig.\til\ref{fig: result LCDM} in the case of an interacting dark sector with kernel given in Section \ref{sec:models} and labeled as case \protect\hyperlink{case I}{I}  (first row), case \protect\hyperlink{case II}{II} (second row), and case \protect\hyperlink{case III}{III}(third row).}
    \label{fig: result interacting }
\end{figure*}

In Figures\til\ref{fig: result LCDM},\til\ref{fig: result interacting }, and we depict the marginalized posterior distributions and the 2D contours in the pair of parameters at the 68\% and 95\% confidence levels with dark and light colors, respectively, and for the three network configurations. Figures depict results obtained for the $\Lambda$CDM and the interacting dark energy models, respectively. The best-fit values of the cosmological parameters with the corresponding statistical uncertainty at 68\% of the confidence level are listed in Table\til\ref{tab: result LCDM} and in Table\til\ref{tab: result Interacting} for the $\Lambda$CDM and the interacting dark energy models, respectively.

\paragraph*{{\bf\emph{{\boldmath$\Lambda$}CDM.}}} 
In Fig.\til\ref{fig: result LCDM}, we depict the marginalized posterior distributions for the pair of parameters ($H_0, \Omega_{\rm m,0})$.  Considering only the ET network, 
the accuracy on the cosmological parameters is: {\em (i)} in the case of  GW+PE $[\sigma_{H_0}, \sigma_{\Omega_{\rm m,0}}]= [3.3,\,0.15]$;   {\em (ii)} in the case of  GW+AF $[\sigma_{H_0}, \sigma_{\Omega_{\rm m,0}}]= [3.4,\, 0.12]$; and  {\em (iii)} in the case of  GW+KN $[\sigma_{H_0}, \sigma_{\Omega_{\rm m,0}}]= [1.4,\,0.25]$.  These results translate into an accuracy on the Hubble constant of $\sim 4.8\%$, $\sim 5\%$ and $\sim 2\%$
and on $\Omega_{\rm m,0}$ of $\sim 48\%$, $\sim 39\%$ and $\sim 78\%$, which is comparable with the values reported in\til\cite{Branchesi:2023mws}. 
Additionally, we found remarkable improvements in the precision of the estimation of the cosmological parameters when ET is used jointly with CE.  Using the network configuration ET+CE, one may constrain the cosmological parameters with the following accuracies: {\em (i)} in the case of  GW+PE $[\sigma_{H_0}, \sigma_{\Omega_{\rm m,0}}]= [1.7,\,0.05]$; {\em (ii)} in the case of  GW+AF $[\sigma_{H_0}, \sigma_{\Omega_{\rm m,0}}]=[2.6,\, 0.09]$; and {\em (iii)} in the case of  GW+KN $[\sigma_{H_0}, \sigma_{\Omega_{\rm m,0}}]=[0.17,\,0.02]$. 
While in the network configuration ET+2CE we find: {\em (i)} in the case of  GW+PE $[\sigma_{H_0}, \sigma_{\Omega_{\rm m,0}}]= [1.1,\,0.04]$; {\em (ii)} in the case of  GW+AF $[\sigma_{H_0}, \sigma_{\Omega_{\rm m,0}}]=[2.4,\, 0.08]$; and {\em (iii)} in the case of  GW+KN $[\sigma_{H_0}, \sigma_{\Omega_{\rm m,0}}]=[0.14,\,0.02]$. Those results show that using ET + 2CE configuration, one may obtain an improvement of a factor of 1.4 and 10 in the cases of GW+AF and GW+KN, respectively, with respect to using only ET. 
These results are also comparable with the constraints presented in\til\cite{Dhani:2022}, which demonstrates the effectiveness of our procedure.

\paragraph*{{\bf\emph{Interacting Scenarios.}}}
In Figure\til\ref{fig: result interacting }, we represent the marginalized posterior distributions corresponding to case \hyperlink{case I}{I}, \hyperlink{case II} and \hyperlink{case III}{III} in the first, second and third rows, respectively.

We show that, in the framework of an interaction kernel $Q$ proportional to the DE density, namely case \hyperlink{case I}{I}, ET will be capable of bounding the cosmological parameters with an accuracy of {\em (i)} $[\sigma_{H_0}, \sigma_{\Omega_{\rm m,0}},\omega,\delta]= [7.0,\,0.33,\,1.8,\,2.6]$ using GW+PE; {\em (ii)} $[\sigma_{H_0}, \sigma_{\Omega_{\rm m,0}},\omega,\delta]= [8.5,\, 0.27,\,1.6,\,1.5]$ using GW+AF; and {\em (iii)} $[\sigma_{H_0}, \sigma_{\Omega_{\rm m,0}},\omega,\delta]= [2.6,\,0.36,\,1.5,\,3.9]$ using GW+KN datasets, respectively. 

On the other hand, if you use the ET + CE network configuration, you can obtain an accuracy of {\em (i)} $[\sigma_{H_0}, \sigma_{\Omega_{\rm m,0}},\omega,\delta]= [9.4,\,0.23,\,1.33,\,1.14]$; {\em (ii)} $[\sigma_{H_0}, \sigma_{\Omega_{\rm m,0}},\omega,\delta]= [7.6,\, 0.44,\,1.3,\,1.2]$; and {\em (iii)} $[\sigma_{H_0}, \sigma_{\Omega_{\rm m,0}},\omega,\delta]= [0.57,\,0.21,\,2.1,\,2.7]$ when using the GW + PE, GW + AF and GW + KN datasets, respectively.  Finally, using the network configuration ET+2CE, the accuracy on cosmological parameters improves of a factor $\sim[1,\,1.6,\,1.5,\,1.2]$ in the case of GW+PE dataset with respect to the ET results. Similar improvements are also found when the  GW+AF dataset is used. On the other hand, for the GW + KN dataset, the accuracy in the cosmological parameters improves by a factor $\sim[9,\,1.7,\,1.5,\,1.4]$, showing a huge improvement in the accuracy of the Hubble constant.
The other two interacting models that we have studied may be constrained with similar accuracies in the cosmological parameters, as shown in Table\til\ref{tab: result Interacting}.  
However, it should be noted that the value of $\Omega_{\rm m,0}$ is never well constrained. Moreover, the parameter w and the interaction parameters $\delta$ or $\eta$ recover the fiducial values at more than the 68\% confidence level, thus showing a small statistical bias.
We argue that these features could be related to the strong degeneracy between the $\Omega_{\rm m,0}$, w and $\delta$ (or $\eta$) as it is clear from equations\til\eqref{eq: E(z) int DE}--\eqref{eq:E(z) int DM}--\eqref{eq:E(z) int total}, and the loss of power constraint due to the fact the in case \hyperlink{case II}{II} and \hyperlink{case III}{III} the $\Omega_{b,0}$ is decoupled from $\Omega_{\rm dm,0}$.
In order to check whether this is the case or not, we repeat the analysis by setting a Gaussian prior on $\Omega_{\rm m,0}$ {with the standard deviation equal to 10\% of the central value. Our  choice is based on the results in\til\cite{Yang:2020uga,Sabogal:2024yha} where interacting DE models have been constrained, and the statistical error on $\Omega_{\rm m,0}$ is roughly 10\%.}
In Table\til\ref{tab: result Interacting DE Om fixed}, we report the best-fist values and the 68\% confidence interval obtained for the case \hyperlink{case I}{I}. Such an assumption helps to eliminate the bias on the  w and $\delta$ parameters recovering the fiducial values at the 68\% confidence level. Therefore, we argue that to use GWs dataset to constrain the interacting models under consideration, one will need to adopt a Gaussian prior on the total matter density parameter to avoid statistical biases.

\begin{table*}[]
    \centering
    \begin{tabular} { c |c| c c c|| c c c|| c c c}
\hline
 & & \multicolumn{3}{c||}{\bf CASE \protect\hyperlink{case I}{I} } & \multicolumn{3}{c||}{\bf CASE \protect\hyperlink{case II}{II}} & \multicolumn{3}{c}{\bf CASE \protect\hyperlink{case III}{III}} \\
\hline
 {\bf }& {\bf Network} & {\bf PE} & {\bf AF} & {\bf KN} & {\bf PE} & {\bf AF} & {\bf KN} & {\bf PE} & {\bf AF} & {\bf KN} \\
\hline
 & ET & $65.5^{+4.1}_{-10.0}$ & $72.3^{+6.1}_{-11.0}$ & $68.2^{+1.8}_{-3.5}$ & $69^{+8}_{-10}$ & $72.1^{+5.6}_{-10.0}$ & $69.3^{+2.4}_{-3.0}$ & $67.7^{+3.8}_{-5.1}$ &$69.9^{+4.1}_{-5.4}$ & $68.5^{+1.7}_{-1.9}$\\
$H_0$ & ET+CE & $83.2^{+8.9}_{-10.0}$ & $72.9^{+5.2}_{-10.0}$ & $67.67^{+0.34}_{-0.40}$ & $78^{+9}_{-7}$ & $71.5^{+5.7}_{-11.0}$ & $67.55^{+0.29}_{-0.36}$ & $71.7^{+3.9}_{-4.6}$ & $68.1^{+3.8}_{-5.1}$ & $67.66^{+0.22}_{-0.38}$\\
 & ET+2CE & $79.2^{+4.6}_{-9.8}$ & $82.7^{+8.0}_{-10.0}$ & $67.68^{+0.27}_{-0.31}$ & $74.6^{+4.8}_{-7.7}$ & $81.5^{+7.9}_{-7.9}$ & $67.58^{+0.24}_{-0.31}$ & $71.3^{+3.1}_{-4.3}$ & $73.4^{+4.7}_{-5.4}$ & $67.65^{+0.18}_{-0.31}$\\
\hline
 & ET & $0.45^{+0.26}_{-0.39}$ & $0.44^{+0.20}_{-0.38}$ & $0.49^{+0.33}_{-0.40}$ & $0.40^{+0.10}_{-0.05}$ & $0.39^{+0.10}_{-0.05}$ & $0.40^{+0.09}_{-0.07}$ & $0.41^{+0.08}_{-0.08}$ & $0.41^{+0.08}_{-0.08}$ & $0.42^{+0.08}_{-0.08}$\\
$\Omega_{\rm m,0}$  & ET+CE & $0.36^{+0.15}_{-0.31}$ & $0.42^{+0.25}_{-0.38}$ & $0.58^{+0.31}_{-0.11}$ & $0.39^{+0.07}_{-0.07}$ & $0.39^{+0.07}_{-0.07}$ & $0.39^{+0.07}_{-0.07}$ & $0.44^{+0.11}_{-0.05}$ & $0.42^{+0.12}_{-0.06}$ & $0.40^{+0.06}_{-0.12}$\\
 & ET+2CE & $0.36^{+0.19}_{-0.32}$ & $0.39^{+0.21}_{-0.21}$ & $0.56^{+0.31}_{-0.11}$ & $0.39^{+0.09}_{-0.05}$ & $0.37^{+0.04}_{-0.09}$ & $0.39^{+0.07}_{-0.07}$ & $0.44^{+0.10}_{-0.05}$ & $0.45^{+0.10}_{-0.05}$ & $0.40^{+0.06}_{-0.12}$\\
\hline
& ET & $-2.7^{+1.5}_{-2.1}$ & $-2.8^{+1.2}_{-2.0}$ & $-2.4^{+1.5}_{-1.5}$ & $-1.6^{+1.5}_{-3.2}$ & $-1.93^{+1.60}_{-0.68}$ & $-1.99^{+1.50}_{-0.84}$ & $-1.6^{+1.9}_{-2.1}$  & $-1.8^{+1.2}_{-1.2}$ & $-1.8^{+1.3}_{-1.3}$\\
w  & ET+CE & $-3.94^{+0.85}_{-1.80}$ & $-2.8^{+1.3}_{-1.3}$ & $-2.43^{+1.80}_{-0.58}$ & $-3.0^{+1.3}_{-2.0}$ & $-2.08^{+1.80}_{-0.91}$ & $-1.11^{+0.25}_{-0.12}$ & $-2.15^{+0.73}_{-1.70}$ & $-1.8^{+1.8}_{-1.6}$ & $-1.01^{+0.79}_{-0.34}$\\
 & ET+2CE & $-3.20^{+0.80}_{-1.60}$ & $-3.0^{+1.2}_{-1.1}$ & $-2.18^{+1.50}_{-0.53}$ & $-2.5^{+1.7}_{-2.4}$ & $-3.0^{+1.1}_{-1.7}$ & $-1.10^{+0.23}_{-0.12}$ & $-2.1^{+1.1}_{-1.1}$  & $-2.21^{+0.73}_{-1.60}$ & $-0.95^{+0.71}_{-0.38}$\\
\hline
& ET & $\phantom{-}2.0^{+1.5}_{-3.8}$ & $\phantom{-}0.1^{+1.3}_{-1.7}$ & $\phantom{-}2.8^{+3.9}_{-3.9}$ & $-0.49^{+0.75}_{-0.28}$ & $-0.41^{+0.61}_{-0.15}$ & $0.21^{+1.70}_{-0.76}$ & $-0.03^{+0.17}_{-0.17}$  & $-0.09^{+0.16}_{-0.16}$ & $\phantom{-}0.12^{+0.38}_{-0.69}$\\
 ${\boldmath\delta \textbf{ or }\eta}$ & ET+CE & $-0.39^{+1.50}_{-0.78}$ & $-0.1^{+1.4}_{-1.0}$ & $-1.29^{+1.90}_{-0.83}$ & $-0.20^{+0.15}_{-0.11}$ & $-0.18^{+0.35}_{-0.07}$ & $-0.42^{+0.83}_{-0.51}$ & $-0.07^{+0.09}_{-0.09}$& $\phantom{-}0.04^{+0.11}_{-0.11}$ & $-0.19^{+0.13}_{-0.24}$\\
 & ET+2CE & $-0.17^{+1.30}_{-0.59}$ & $-0.69^{+1.30}_{-0.58}$ & $-1.09^{+1.60}_{-0.72}$ & $-0.16^{+0.21}_{-0.21}$ & $-0.27^{+0.19}_{-0.19}$ & $-0.42^{+0.74}_{-0.43}$ & $-0.05^{+0.08}_{-0.08}$  & $-0.14^{+0.10}_{-0.10}$ & $-0.21^{+0.12}_{-0.23}$\\
\hline
\end{tabular}
    \caption{The same of Table\til\ref{tab: result LCDM} in the case of an interacting dark sector with kernel given in Section \ref{sec:models} and labeled as case \protect\hyperlink{case I}{I}, case \protect\hyperlink{case II}{II}, and case \protect\hyperlink{case III}{III}.}
    \label{tab: result Interacting}
\end{table*}

\begin{table}[]
    \centering
    \begin{tabular} { c |c| c c c}
    \hline
 {\bf Parameter} &  {\bf Network} & {\bf PE} &  {  \bf AF} & {\bf KN}\\
\hline 
&ET& $71.2^{+5.6}_{-11.0}         $ & $75.4^{+7.3}_{-11.0}         $ & $69.1^{+2.3}_{-2.7}        $\\
{\boldmath$H_0$}&ET+ CE& $83^{+9}_{-7}             $ & $75.1^{+6.9}_{-9.6}        $ & $67.69^{+0.34}_{-0.40}     $\\
 & ET+2CE & $74.9^{+3.0}_{-7.9}        $ & $85^{+10}_{-7}             $ & $67.71^{+0.28}_{-0.32}     $\\
\hline
&ET& $0.311\pm 0.031            $ & $0.311\pm 0.031            $ & $0.312\pm 0.031            $\\

{\boldmath$\Omega_{\rm m,0}   $} &ET+ CE & $0.310\pm 0.031            $ & $0.312\pm 0.032            $ & $0.313\pm 0.031            $\\

& ET+2CE& $0.312\pm 0.031            $ & $0.312\pm 0.031            $ & $0.312\pm 0.032            $\\
\hline
&ET & $-1.7^{+1.4}_{-1.4}              $ & $-1.7^{+1.8}_{-1.2}        $ & $-1.90^{+1.70}_{-0.77}      $\\

{\boldmath$\omega   $} &ET+ CE& $-1.11^{+0.75}_{-1.80}      $ & $-1.6^{+1.7}_{-1.1}        $ & $-1.05^{+0.19}_{-0.15}     $\\

& ET+2CE& $-1.02^{+1.20}_{-0.39}      $ & $-1.81^{+0.96}_{-0.96}             $ & $-1.04^{+0.16}_{-0.12}     $\\
\hline
&ET & $1.68^{+0.26}_{-2.10}       $ & $0.59^{+0.12}_{-0.91}      $ & $1.7^{+3.5}_{-1.6}         $\\
{\boldmath$\delta   $} &ET+ CE& $-0.06^{+0.30}_{-0.22}     $ & $0.47^{+0.21}_{-0.60}      $ & $0.11^{+0.42}_{-0.42}$\\
& ET+2CE  & $0.04^{+0.17}_{-0.17}             $ & $-0.26^{+0.28}_{-0.21}     $ & $0.09^{+0.34}_{-0.34}           $\\
\hline
\end{tabular}
    \caption{Cosmological inference of the parameters of the interacting dark sector with kernel labeled as case \protect\hyperlink{case I}{I}, and obtained by adopting a Gaussian prior on $\Omega_{\rm m,0}.$}
    \label{tab: result Interacting DE Om fixed}
\end{table}

\section{Conclusion}\label{sec:conclusion}

The detection of gravitational waves has opened up new ways to explore the Universe. In the near future, 3G GW detectors will be real sources of discoveries that will shed light on open questions in modern cosmology, such as dark matter and dark energy. Among the many possibilities, on the test bench will be models in which these two exotic supplies of matter and energy can interact with each other, impacting the dynamics of the Universe.

We constructed a catalog of GW events from BNS to explore these models and understand the possibilities of verifying or rejecting them. To build this catalog, we used the sensitivity curves of future observatories such as ET and CE, and we have determined how many GW events can be detected yearly, and their corresponding redshift and SNR. The statistical error on the model parameters of interest has been determined using the FIM. Finally, we also considered the possibility that a GRB or a KN event, linked with GW events, can be detected by X-ray and optical observatories such as THESEUS and VRO, respectively.

We explored three different DE models in which the interaction is proportional to DE, DM and the dark sector's energy density. Our analysis is based on five years of observations and carried out using a MCMC algorithm. The results show that the Hubble constant can be at most constrained with a 10\% accuracy, which is at least a factor 10 higher than the one required to solve the Hubble tension \cite{Califano:2023fbq}. Things get worse when one looks at $\Omega_{\rm m,0}$, which is poorly constrained. The same happens for the parameters of the interaction kernel. The main source of such poor results is the strong degeneracy between $\Omega_{\rm m,0}$ and the parameters of the interaction kernel. Indeed, by adopting a Gaussian prior on $\Omega_{\rm m,0}$, one can break the degeneracy and allow to constrain the parameters of the interaction kernel. However, the precision in $H_0$ remains at $\sim 10\%$ making troubles to resolve the Hubble tension in the framework of interacting DE models.

\section*{Acknowledgments}
MC acknowledges the support of Istituto Nazionale di Fisica Nucleare (INFN) {\it iniziativa specifica} MOONLIGHT2. IDM acknowledges financial support from the grant PID2021-122938NB-I00 funded by MCIN/AEI/10.13039/501100011033 and by “ERDF A way of making Europe”, and from the grant SA097P24 funded by Junta
de Castilla y León.
\clearpage

\bibliographystyle{apsrev4-2}
\bibliography{Biblio.bib}

\end{document}